\documentclass[aps, oneprint, twoside, superscriptaddress, nofootinbib,
floatfix, notitlepage, onecolumn, 11pt, tightenlines]{revtex4-1}

\usepackage{amsmath,amsthm}
\usepackage{amssymb,enumerate}
\usepackage{bm}
\usepackage[dvipsnames]{xcolor}
\usepackage{float}
\usepackage[T1]{fontenc}
\usepackage[pdftex]{graphicx}
\usepackage[pdftex,colorlinks=true,linkcolor=blue,citecolor=blue,urlcolor=blue]{hyperref}
\usepackage{comment}
\usepackage[caption=false]{subfig}
\usepackage{multirow}
\usepackage{tabularx, booktabs}
\usepackage[normalem]{ulem}
\usepackage{yquant}
\useyquantlanguage{groups}
\usepackage{bm}
\usepackage[margin=1.1in]{geometry}
\usepackage{hyperref}

\usepackage{tikz}

\newcommand{\bra}[1]{\langle #1 |}
\newcommand{\ket}[1]{| #1 \rangle}
\newcommand{\braket}[1]{\langle #1 \rangle}
\newcommand{\mean}[1]{\langle #1 \rangle}
\newcommand{\SK}{\textnormal{SK}}

\newcommand{\vect}[1]{\boldsymbol{#1}}
\newcommand{\gammav}{\vect{\gamma}}
\newcommand{\betav}{\vect{\beta}}
\newcommand{\R}{\mathbb{R}}
\newcommand{\Z}{\mathbb{Z}}

\newcolumntype{Y}{>{\centering\arraybackslash}X}

\begin{document}

\title{Quantum-Enhanced Optimization by Warm Starts}
\author{Ieva \v{C}epait\.{e}}
\affiliation{Phasecraft Ltd.}
\author{Niam Vaishnav}
\affiliation{Phasecraft Ltd.}
\author{Leo Zhou}
\affiliation{Phasecraft Ltd.}
\affiliation{University of California, Los Angeles}
\author{Ashley Montanaro}
\affiliation{Phasecraft Ltd.}
\affiliation{University of Bristol}

\date{\today}

\begin{abstract}
We present an approach, which we term quantum-enhanced optimization, to accelerate classical optimization algorithms by leveraging quantum sampling. Our method uses quantum-generated samples as warm starts to classical heuristics for solving challenging combinatorial problems like Max-Cut and Maximum Independent Set (MIS). To implement the method efficiently, we introduce novel parameter-setting strategies for the Quantum Approximate Optimisation Algorithm (QAOA), qubit mapping and routing techniques to reduce gate counts, and error-mitigation techniques. Experimental results, including on quantum hardware, showcase runtime improvements compared with the original classical algorithms.
\end{abstract}

\maketitle

\section{Introduction}
\label{sec:introduction}

Solving hard optimization and constraint satisfaction problems has long been predicted to be a significant application of quantum computers. These problems, which are ubiquitous in industries such as logistics, finance, telecommunications and healthcare, are crucial for developing efficient processes, reducing costs, and improving decision making. 

Traditional quantum approaches -- like Grover's famous quantum unstructured search algorithm~\cite{grover97} -- can achieve rigorous speedups over corresponding classical search algorithms. However, they require deep quantum circuits and hence inevitably necessitate fault tolerance. In addition, the quantum speedups obtained are usually at most quadratic, and the overheads associated with fault-tolerance may reduce or even eliminate these speedups in practice~\cite{campbell19,babbush21}. 

The Quantum Approximate Optimization Algorithm~\cite{1411.4028} (QAOA) represents a promising alternative that may be more suitable for near-term quantum computers.
Unlike Grover’s algorithm, the focus of QAOA is the minimization of a cost function constructed as a sum of local costs. QAOA takes problem structure into account and there is some evidence that it can outperform classical algorithms~\cite{boulebnane24,farhi19, basso_et_al:LIPIcs.TQC.2022.7, montanaro2024symmetric}, perhaps even exponentially in some cases. In addition, the simple structure of the QAOA algorithm means that, for certain problems, the algorithm can easily be implemented directly on a NISQ quantum computer\footnote{Quantum computers today are said to be in the NISQ (Noisy Intermediate Scale Quantum) era~\cite{preskill18}. NISQ quantum computers with physically reasonable error rates have at most hundreds of qubits, and can execute at most thousands of 2-qubit gates.}, with problem variables mapping directly to qubits, and the problem structure mapping to quantum gates.

Despite these advantages, the limited capabilities of today’s quantum hardware and the significant advancements that have been made in classical optimization algorithms pose challenges for QAOA to outperform the best classical approaches directly. Classical optimization algorithms for large-scale combinatorial optimization can be divided into two broad categories: exact algorithms based on techniques such as branch-and-bound~\cite{gurobi, automatic1960land}, which guarantee that an optimal solution can be found (but which may take exponential time), and heuristics~\cite{burer_rank-two_2002, palubeckis_multistart_2004, festa_randomized_2002}, which make no such guarantees, but can run much more quickly. State-of-the-art exact optimization algorithms can solve arbitrary instances of many important optimization problems -- such as the Max-Cut problem, which we will discuss below -- on up to about 100 variables~\cite{gurobi}, depending on the structure of the problem. State-of-the-art heuristics can solve instances on hundreds or thousands of variables to optimality in practice~\cite{dunning_what_2018}.
This raises the question of whether NISQ-era quantum computers could be of any use in solving hard optimization problems, or if we will need to wait until the availability of large-scale fault-tolerant quantum computers, with thousands of logical qubits and the ability to execute millions of quantum gates.

\subsection{Our approach}

Here we introduce an approach which we term \emph{quantum-enhanced optimization}, where noisy samples from a near-term quantum algorithm such as QAOA are used as a ``warm start'' to accelerate classical optimization algorithms. This could enable a quantum speedup to be obtained before quantum computers become competitive with classical ones at solving the entirety of a hard instance directly. 
By leveraging the strengths of both classical and quantum computing, our approach aims to take advantage of the potential of NISQ devices in order to enhance classical optimization methods, making it possible to tackle complex optimization challenges more effectively.

In our approach, we use a quantum algorithm (in particular, the QAOA algorithm, although any other quantum algorithm could be used) to generate samples that are used as input into a classical algorithm. The procedure is as follows:
\begin{enumerate}
\item The overall optimization problem is divided into small subproblems;
\item Each subproblem is approximately solved efficiently using a quantum computer, which produces a list of relatively low-cost solutions;
\item These low-cost solutions are combined to produce a list of potential solutions to the overall problem;
\item These solutions are used as starting points (``warm starts'') for a classical heuristic;
\item The classical heuristic finds a good solution to the overall problem, ideally an optimal one, more quickly than it would have done without the warm start.
\end{enumerate}
Each of these steps involves a number of technical decisions and the implementation of several underlying technical components to ensure high-quality results.
Quantum-enhanced optimization has a number of appealing properties:

\begin{enumerate}
    \item Any classical heuristic that relies on being seeded with a random bit-string has potential to be accelerated via our approach. For example, many algorithms start with a uniformly distributed bit-string, then attempt to improve on this via some kind of search procedure. Categories of algorithms where our approach may work include, but are not limited to, \emph{local search algorithms}, \emph{genetic algorithms} and \emph{simulated annealing}. Further, to get an advantage over the original classical algorithm, the quantum algorithm only needs to produce a more useful input than a uniformly random bit-string.
    \item The quantum-enhanced algorithm should degrade gracefully with noise. In the case of completely catastrophic depolarizing noise, the output of the quantum part is simply a uniformly random bit-string, which is what the classical algorithm used in the first place. This property need not hold for biased noise models. However, for any type of noise, one can also view the classical postprocessing as an error-correction procedure that will help to correct the noise.
    \item The quantum computer is only used once, at the start of the algorithm, and samples can be stored offline and used later.
    \item Because the quantum computer is used only to generate classical samples, a large problem can be divided into smaller subproblems, which can be solved and the solutions recombined.
    \item The quantum computer is used for sampling, rather than measuring observables. This is harder to simulate using classical methods, and gives hope that we can produce distributions with interesting correlations. Indeed, generating samples from even constant-depth noiseless QAOA is hard for classical algorithms~\cite{farhi19} (given some reasonable computational complexity assumptions), whereas expectation values of local observables can be computed in time $O(1)$. While recent approaches have been proposed for classical noisy sampling from QAOA based on approximating expectation values~\cite{martinez2025sampling}, the practicality of such methods with regards to quality of samples versus efficiency of their generation, has not yet fully been explored.
\end{enumerate}


\subsection{Summary of results}

\begin{figure}
\includegraphics[width=\textwidth]{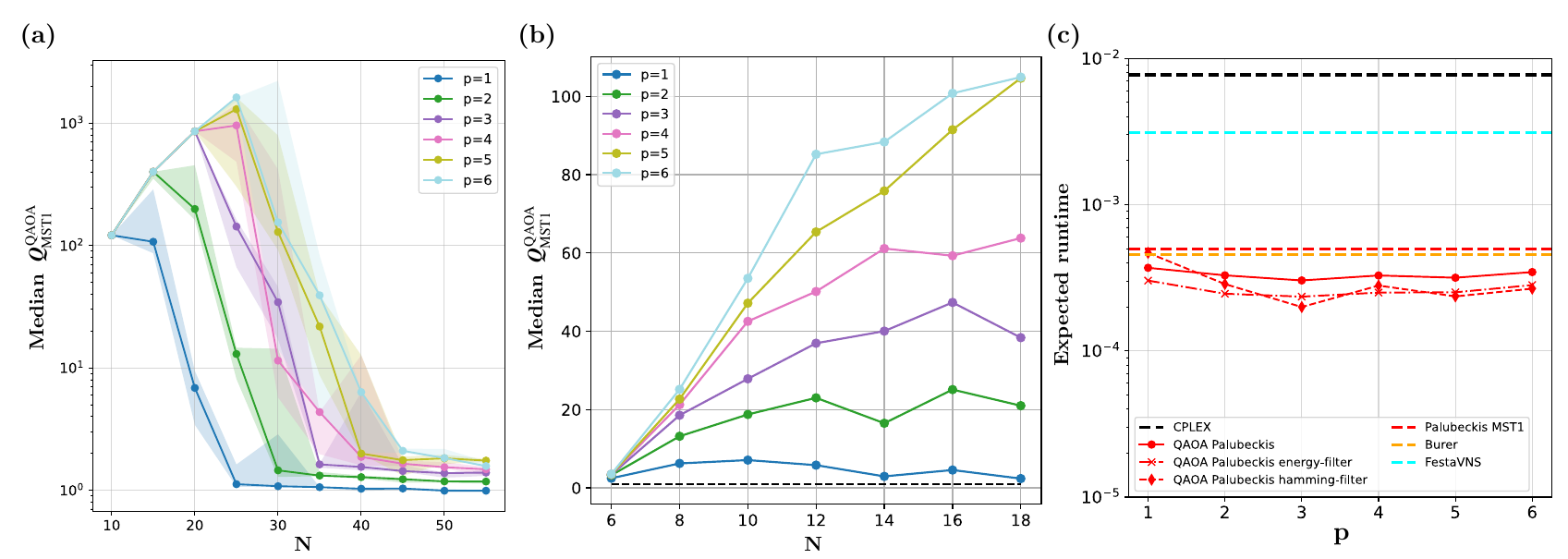}
    \caption{Summary of key numerical results. (a) Speed-up in solving Max-Cut on line graph problem instances using QAOA samples from IBM hardware, where $N$ is the length of the line. Plot demonstrates the Q-factor ($Q_{\rm MST1}^{\rm QAOA}$, Eq.~\eqref{eq:qfactor}) over a classical tabu search heuristic algorithm PalubeckisMST1~\cite{palubeckis_multistart_2004}, when the random initial bit-strings are replaced by QAOA samples obtained from circuits runs with $p$ layers, and post-processed with a method we refer to as the `hamming-filter'. The line plot is a median across 3 separate sets of samples obtained from 3 hardware runs, while the shaded region encompasses the min and max speed-up obtained. For more details see Sec.~\ref{sec:maxcut_line_graphs}. (b) Speed-up factor over classical tabu search for degree 4 random regular graphs when using QAOA samples obtained from emulations of the corresponding circuits. For more details see Sec.~\ref{sec:emulated_results}. (c) Median expected runtime (Eq.~\eqref{eq:expected_runtime}) results across 10 instances of weighted Max-Cut with weights randomly chosen in the range $[-1,1]$ on a 41 vertex graph, where each horizontal line corresponds to a classical solver, while the quantum-enhanced results are plotted for each number of QAOA layers $p$. The classical algorithms include CPLEX~\cite{cplex2009v12}, an exact solver, as well as FestaVNS~\cite{festa_2002}, Burer~\cite{burer_rank-two_2002} and PalubeckisMST1~\cite{palubeckis_multistart_2004}. The quantum-enhanced results are obtained by seeding the PalubeckisMST1 algorithm with quantum samples obtained from hardware runs, along with different post-processing techniques. For more details see Sec.~\ref{sec:enhanced_ibm_hw_results} and Sec.~\ref{sec:em_filters}.}
    \label{fig:summary}
\end{figure}

We implemented a full software pipeline that instantiates the approach above, and tested its performance in emulation and on IBM quantum hardware. This software stack is \href{https://optimization.phasecraft.io}{available online}. To obtain good performance, we developed the following new technical ingredients:
\begin{enumerate}
    \item \textbf{Approaches to efficiently choosing QAOA parameters without needing to optimize them variationally.} A key aspect of quantum-enhanced optimization is that the quantum component should run quickly -- using the quantum computer adds an overhead when compared to running solely the original classical optimization heuristic, and if this overhead is too large, any quantum speedup would be lost. If the variational optimization part of QAOA is removed, the quantum computer only needs to be accessed once, at the start of the algorithm. Here we introduce new methods for choosing good QAOA parameters for Max-Cut and Maximum Independent Set (MIS), which are two well-studied NP-hard graph problems, and also more general Quadratic Unconstrained Binary Optimization (QUBO) problems. These approaches are detailed in Sec.~\ref{sec:parameters}.

    \item \textbf{New methods for implementing the QAOA algorithm on quantum hardware.} Although the QAOA algorithm has a very simple structure, it can still be challenging to implement the required gates for a given hardware connectivity. Here we develop two new solutions to this ``mapping and routing'' problem on quantum hardware, one of which is a greedy approach based on a variant of a method used in the context of fermionic swap networks~\cite{clinton24}, the other of which is based on the A* algorithm~\cite{hart_a_star}. Our experiments show an average reduction in number of swaps of 19.1\% and 22.0\% compared with the previously developed 2QAN algorithm~\cite{lao22} and LightSABRE~\cite{zou24_sabre} algorithm respectively. We present our new methods in Sec.~\ref{sec:hardware} and their numerical evaluation in Sec.~\ref{sec:circuit_optimisation_results}.

    \item \textbf{New error mitigation methods.} To achieve high-quality experimental results on quantum hardware requires error mitigation. However, many standard error mitigation methods used in the literature cannot be used in our context, as we use the quantum computer for sampling bit-strings, rather than computing observables. As well as using existing techniques for readout error mitigation, we introduce methods for filtering the set of samples retrieved based on frequency and/or energy. The details of these approaches can be found in Sec.~\ref{sec:error_mitigation}.
\end{enumerate}
We evaluate our approach via a metric which we term the ``Q-factor'', which measures the ratio between the running time of the original classical algorithm and the quantum-enhanced classical algorithm. We find that, both in experiments on quantum hardware and in emulation, in some cases, the quantum-enhanced running time is up to a factor of $\sim1000$ lower than the original classical running time. In addition, the running time improvement increases with problem size, up to instances on 40 qubits on quantum hardware. Some of our key results are summarized in Figure \ref{fig:summary}, with full details in Section \ref{sec:results}.

We note that in order to solve problems which are larger than those which could be fit on current quantum devices, our approach is amenable to the partitioning of large-scale optimization instances prior to running each partition on quantum hardware. We discuss this component further in Sec.~\ref{sec:division}. For the remainder of this document, when describing our techniques and results, for simplicity we will consider the case where the overall problem we are solving does not need to be split up and recombined. However, all of our methods apply in the more general case where the quantum algorithm produces solutions to subproblems that are then combined, and we have implemented this approach in our software stack.
\subsection{Related work}

Another setting in which the notion of ``warm starting'' a classical algorithm with the output of a quantum algorithm has been used is accelerating variational Monte Carlo methods using states produced by the VQE algorithm~\cite{montanaro23}. Our work can be seen as the special case of this approach where the VQE algorithm is applied to a classical Hamiltonian. Our approach should not be confused with the complementary concept of warm-starting QAOA with a state based on the output of a classical algorithm~\cite{egger21,tate23,tate232}. In Ref.~\cite{tabu2021vedran}, the authors use a similar approach of exploiting QAOA outputs to help a classical solver, however in this case they only use it to decide the value of a few key variables in a much larger problem space.

\subsection{Organisation}

We begin in Section \ref{sec:background} by introducing the QAOA algorithm and the optimization problems that we will address. In Section \ref{sec:theory} we define the metric which we will use to measure quantum speedup. Sections \ref{sec:parameters}, \ref{sec:hardware}, \ref{sec:error_mitigation} and \ref{sec:division} describe our techniques: parameter prediction, qubit mapping and routing, error mitigation, and division into smaller instances for larger instances, respectively. Section \ref{sec:results} describes our numerical and experimental results. Finally, Section \ref{sec:discussion} concludes with a summary of our results and discussion of future work. Some additional results are deferred to appendices, in particular results on additional families of graphs, and on accelerating the FESTA2002VNSPR~\cite{dunning_what_2018, festa_2002} algorithm.

\section{Background}
\label{sec:background}

\subsection{The Quantum Approximate Optimization Algorithm}

The Quantum Approximate Optimization Algorithm (QAOA), introduced by Farhi et al.~\cite{1411.4028}, is a variational quantum algorithm for addressing combinatorial optimization or constraint satisfaction problems\footnote{An essentially identical algorithm, but without the variational optimization aspect, was previously proposed by Hogg~\cite{hogg00}.}. This variational circuit consists of a certain number of layers, commonly denoted by $p$. Each layer depends on two variational parameters, also known as \textit{QAOA angles}, which require optimization:
\begin{align}
\label{eq:qaoa_angles}
    \hspace{8em} \gamma_k, \beta_k \in \mathbb{R}, \hspace{3em} \text{where } k \in \{1, ..., p\}.
\end{align}
QAOA aims to find a bit-string $\bm{x} \in \{0, 1\}^n$ minimizing a classical cost function
\begin{align}
    C\left(\bm{x}\right), \qquad \bm{x} \in \{0, 1\}^n.
\end{align}
It does so by preparing a variational quantum state which can then be measured in the computational basis, providing a set of bit-strings which act as candidate solutions to the overarching combinatorial optimization problem. In general, given a classical cost function $C$ of an $n$-bit bit-string, QAOA with $p$ layers prepares the following variational state:
\begin{align}
    \ket{\Psi(\gammav, \betav)} & = \exp\left(i \beta_p H_B\right) \exp\left(i \gamma_p H_C\right) \left(\cdots\right) \exp\left(i \beta_1 H_B\right) \exp\left(i \gamma_1 H_C\right)\ket{+}^{\otimes n},
\end{align}
where
\begin{align}
    H_B & = \sum_{j=1}^{n}X_j
\end{align}
is a transverse field Hamiltonian and
\begin{align}\label{eq:cost_hamiltonian}
    H_C & = \sum_{\bm{x} \in \{0, 1\}^n}C(\bm{x})\ket{\bm{x}}\bra{\bm{x}}
\end{align}
is the Hamiltonian diagonal in the computational basis corresponding to cost function $C$. The variational nature of the QAOA algorithm means that it is not fully specified, requiring an optimization procedure to determine the optimal values of $\gamma$ and $\beta$ which lead to the best (generally lowest cost) output bit-string distributions. This can be done via an online optimization using a quantum device, an offline optimization using an emulator in cases where this is possible, or via some more complex analytic parameter-setting techniques such as those discussed in Sec.~\ref{sec:parameters}.

\subsection{Quadratic Unconstrained Binary Optimization}


The cost Hamiltonian in QAOA can be implemented particularly straightforwardly for a class of problems known as Quadratic Unconstrained Binary Optimization (QUBO). A QUBO problem is defined as:
\begin{align}\label{eq:qubo}
    \text{max} \quad x^T Q x + c^T x& \\
    x \in \{0,1\}^n \quad Q \in \mathbb{R}^{n \times n} \quad c \in \mathbb{R}^n& 
\end{align}
QUBOs can be used to express many well-known NP-complete problems over binary variables given a particular choice of $Q$ and $c$~\cite{lucas_ising_2014}. In this work we will primarily focus on Max-Cut and Maximum Independent Set, two well-studied graph-theoretic problems in the context of QAOA, defined by a particular structure of $Q$ and $c$. The relationship between a QUBO formulation and a spin Hamiltonian whose ground state is the solution to the QUBO is provided via a simple mapping between the binary variables $x$ and the expectation values of quantum operators $Z$.

\paragraph{Weighted Max-Cut}

The goal of Max-Cut is to find a partition of the vertices in a graph into two sets such that the number of edges between the sets is maximised. Formally, for a graph $G(V,E)$ with $|V| = n$ and weights $w_{ij} \in \R$ for $(i,j) \in E$, where $w_{ij} \geq 0$, the goal is to compute
\[ \min_{z \in \{\pm1\}^n} \sum_{(i,j) \in E} w_{ij} z_i z_j. \]
Note that this is a more general version of the commonly stated Max-Cut problem wherein the weights $w_{ij}$ are all set to $1$. This immediately corresponds to a cost Hamiltonian
\begin{align}\label{eq:maxcut_hamiltonian}
    H_{\rm MC} = \sum_{(i,j) \in E} w_{ij} Z_i Z_j.
\end{align}
In the QUBO formulation, we perform a change of variables and this problem becomes
\[
\max_{x \in \{0,1\}^n} \sum_{(i,j) \in E} w_{ij} \left( x_i + x_j - 2x_i x_j \right) 
\]
where the coefficients of the quadratic and linear terms in $x$ are captured by $Q$ and $c$ in the QUBO formulation (Eq.~\eqref{eq:qubo}).

\paragraph{Maximum Independent Set (MIS)}

The goal of the MIS problem for a given graph is instead to determine its largest possible independent set -- one where no two vertices in the set are connected by an edge. The MIS problem can be formally stated as follows:
\begin{align}
    &\max_{x \in \{0,1\}^n} \sum_{i \in V} x_i \\
    \text{s.t. } & \forall (i,j) \in E, \quad x_i + x_j < 1.
\end{align}
This can then be expressed in the standard QUBO formulation by changing it to a maximisation problem with an added Lagrange multiplier $\lambda$:
\begin{align}\label{eq:mis_qubo}
    \max_{x \in \{0,1\}^n} \left(\sum_{i \in V} x_i - \lambda \sum_{(i,j)\in E} x_i x_j,\right)
\end{align}
where $\lambda \geq 1$ is a penalty term that can be modified in order to change the optimization landscape of the problem instance. The cost Hamiltonian for this problem can be formulated as:
\begin{align}\label{eq:mis_hamiltonian}
   H_{\rm MIS} = \sum_{i \in V} \left( \frac{1}{2} - \frac{\lambda d_i}{4} \right) Z_i + \frac{\lambda}{4} \sum_{(i,j) \in E} Z_i Z_j,
\end{align}
where $d_i$ is the degree of the $i^{\rm th}$ vertex. 

\subsection{Reductions to other graph-theoretic problems and applications to power networks}

The Max-Cut problem has many direct applications in a real-world setting, including image segmentation~\cite{desousa13}, declustering of files~\cite{liu96}, half-duplex communication networks~\cite{meshnetworks}, and circuit layout design~\cite{application1988barahona}, among many others. In addition, it turns out that many other fundamental problems in graph theory can easily be reduced to this problem. Max-Cut is NP-complete, so a vast number of other problems can be reduced to it in principle -- but for some graph-theoretic problems, the reduction is particularly simple and lends itself to implementation on quantum hardware. MIS has a range of applications in numerous fields~\cite{butenko_maximum_2003, wurtz_industry_2022}, including antenna placement strategies which maximize coverage given distance constraints~\cite{wurtz_industry_2022}, financial portfolio selection~\cite{hidaka23} and ad-hoc networks~\cite{WU20061}. Finally, the QUBO problem corresponds directly to the problem of the `Ising spin glass', and covers a broad range of NP problem formulations and applications~\cite{lucas_ising_2014}.

\begin{table}[t]
\begin{tabularx}{\textwidth}{XXX}
\toprule
Power network problem & Graph-theoretic problem & Relation to Max-Cut/QUBO \\
\midrule
 Maximal power section~\cite{Jing2023} & Max-Cut  & Direct application of Max-Cut \\
 Optimal islanding scheme~\cite{WANG2023108857} & Minimal balanced cut & Modification of Max-Cut \\
 Critical node detection~\cite{LALOU201892} & Maximum Independent Set & QUBO formulation (Eq.~\eqref{eq:mis_qubo}) \\
 PMU placement~\cite{brueni_pmu_2005} & Minimal vertex cover& Conjugate problem to the MIS \\ \bottomrule
\end{tabularx}
\caption{Relationship between Max-Cut/QUBO formulations of NP graph problems and power network applications.}
\label{tab:powergrid_applications}
\end{table}

Other graph problems which can be formulated easily via either a Max-Cut reduction or a QUBO include the \emph{minimum balanced cut} and \emph{minimum vertex cover}. One area where all of these problems turn out to occur very naturally is power network design and optimization. These problems are summarised in Table~\ref{tab:powergrid_applications}.

\section{Accelerating classical heuristics}
\label{sec:theory}

Our approach can be applied to accelerate many classical heuristic algorithms for QUBO problems. The work in Ref.~\cite{dunning_what_2018} offers a systematic review of many state-of-the-art heuristics from all three key categories of algorithms that may be accelerated by our method: local search, genetic algorithms and simulated annealing. The authors evaluate the performance of each across several metrics using a custom library of problem instances. In this work, we will largely focus on accelerating an algorithm from this review called ``PalubeckisMST1'', which was originally published in \cite{palubeckis_multistart_2004}. This algorithm, and its variations, is one of the highest-performing heuristics for Max-Cut in the comprehensive analysis given in \cite{dunning_what_2018} (Table \ref{tab:mqlib_warmstarts}) and is based on the use of a uniformly random bit-string as input. Its choice as a benchmark is further motivated by the fact that its core is a local search heuristic `tabu search', which is an important subroutine of several other, high-performing algorithms in the review. We also demonstrate acceleration of the ``FESTA2002VNSPR'' algorithm~\cite{festa_2002}, and defer these results to Appendix~\ref{sec:festa_results}.

We will consider various metrics with which to measure the level of speedup achieved by our approach.
Many heuristic algorithms, such as PalubeckisMST1, have the following structure:
\begin{enumerate}
\item Repeat the following procedure $R$ times:
\begin{enumerate}
    \item Generate a random bit-string $s$;
    \item Perform a search procedure starting from $s$, and stop after $T$ iterations.
\end{enumerate}
\end{enumerate}
A metric to determine the efficiency of different sets of bit-strings for initialising a heuristic is an inspection of the probability $F_{\rm opt}(T)$ that the algorithm returns an optimal solution within a given number of iterations $T$. Ideally, one would hope to minimise $T$ while simultaneously maximising the probability of the heuristic returning an optimal solution. This leads to a metric or `cost' associated with the limit $T$ on the number of iterations that the algorithm can run for:
\begin{align}
    C(T) = \frac{T}{F_{\text{opt}}(T)},
\end{align}
where $T$ is the number of iterations and $F_{\text{opt}}(T)$ is the probability of returning an optimal solution after $T$ iterations. This metric is equivalent to the time-to-solution (TTS) metric~\cite{ronnow2014defining} often used for evaluating optimization runtime when $F_{\text{opt}}$ is small. As the expected number of repetitions required to find an optimal solution is $1/F_{\text{opt}}(T)$, and each repetition uses $T$ iterations, $C(T)$ is the expected number of iterations required to find an optimal solution. Minimising $C(T)$ over $T$ makes the algorithm find an optimal solution as quickly as possible. In practice, $F_{\rm opt}(T)$ can be estimated as the fraction of optimal solutions found, taken over many repetitions of the algorithm. This gives a notion of `minimal expected runtime' of the algorithm:
\begin{table}[t]
\begin{tabularx}{\textwidth}{XXXX}
\toprule
Algorithm name & Algorithm type & Problem type & Average rank in \cite{dunning_what_2018} \\
\midrule
PalubeckisMST1~\cite{palubeckis_multistart_2004} & Local search (tabu) & QUBO & 15.2  \\
Festa02 and its variants~\cite{festa_randomized_2002} & Local search (variable neighborhood search) & Max-Cut &  13.8 \\
Lu10~\cite{lu_hybrid_2010} & Genetic algorithm with local search (tabu) & QUBO & 13.1 \\
Beasley1998SA~\cite{beasley_heuristic_1998} & Simulated Annealing & QUBO & 30.8 \\ \bottomrule
\end{tabularx}
\caption{A selection of algorithms listed in \cite{dunning_what_2018} which could be accelerated with QAOA. Each was designed with either the Max-Cut or QUBO problem in mind, but as a simple transformation exists between the two problems, either can be addressed using any of the algorithms in the list. We provide some examples of runtime estimates for the algorithms from Ref.~\cite{dunning_what_2018} in Appendix~\ref{sec:classical_benchmarks}.}
\label{tab:mqlib_warmstarts}
\end{table}
\begin{align}\label{eq:expected_runtime}
    R_{\min} = \min_T C(T).
\end{align}
When initialized with QAOA-generated samples, the cost is expected to decrease. Thus, for the same problem instance, we expect the minimum of $C(T)$, i.e.~the expected runtime, to be lower for low-cost initial bit-strings when compared to a random initialization, leading to a more efficient performance of the heuristic. A good metric to quantify this efficiency improvement or `speed-up', is to compute the ratio of the two minima obtained using different initialisation strategies:
\begin{align}\label{eq:qfactor}
    Q = \frac{\min_T C_{\text{random}}(T)}{\min_T C_{\text{warm-start}}(T)},
\end{align}
where $Q > 1$ indicates a speedup, e.g.~$Q = 2$ implies that the warm starts made the algorithm converge twice as quickly as the case where random bit-strings were used. We will refer to this metric as the \emph{Q-factor}. In practice, we can compute the Q-factor by running a large number of optimizations for the same problem instance with a large limit set for the total number of iterations $T_{\rm total}$ using either random initialisation or QAOA warm starts and then using those statistics to obtain $F_{\text{opt}}(T)$ for all $T < T_{\rm total}$.

\section{Parameter-setting techniques}
\label{sec:parameters}

One of the key factors in the performance of QAOA is the choice of parameters $\gammav$ and $\betav$ at each layer. In the simplest case, when $p = 1$, we can compute the output state energies efficiently using an analytic approach \cite{Ozaeta_2022}, bypassing the need for parameter-prediction. However, for $p > 1$, this is no longer the case, and more involved methods become necessary, as will be discussed below.

In order to evaluate the performance of our prediction methods, we will use a rescaled approximation ratio metric:
\begin{align}\label{eq:approx_ratio_rescaled}
    \text{AR*}(\bm{\gamma}, \bm{\beta}) = 1 - \frac{ E_{\rm p, QAOA}(\bm{\gamma}, \bm{\beta})  - \lambda_{\rm min}}{\lambda_{\rm max} - \lambda_{\rm min}},
\end{align}
where $E_{\rm p, QAOA}(\bm{\gamma}, \bm{\beta})$ is the energy of a state obtained using QAOA with $p$ layers and some parameters $(\bm{\gamma}, \bm{\beta})$, which were computed via a prediction method, or classical optimization. The parameters $\lambda_{\rm min}, \lambda_{\rm max}$ are the minimum and maximum eigenvalue of the cost Hamiltonian respectively, for a particular QUBO instance. Thus, the AR* metric is $1$ when the obtained QAOA state is the ground state of $H_C$ (Eq.~\eqref{eq:cost_hamiltonian}) and $0$ when it is the maximally excited state. This metric is particularly useful for our purposes as it allows us to treat different problem sizes and problem instance structures on the same footing, by rescaling the full eigenspectrum of each problem Hamiltonian to be comparable. As we wish to explore how the prediction methods fare across all types of problems, the AR* allows us to stick to a single, informative metric.

When it comes to the classical optimization of QAOA parameters, we implement a bounded \texttt{L-BFGS-B} algorithm~\cite{lbfgsb} with 5 separate initializations for each problem instance, using both random as well as informed linear-ramp-like initial parameter guesses. While it is possible to obtain better optimization results by e.g.~performing the optimization more times or using more informed initial guesses, we wanted to demonstrate a bare-bones comparison, as the prediction methods do not require numerous refinements in order to produce the parameters.

\subsection{Weighted Max-Cut}\label{sec:maxcut_angle_prediction}

The weighted Max-Cut Hamiltonian of Eq.~\eqref{eq:maxcut_hamiltonian} is made up of only $ZZ$ interactions corresponding to the edges of a given problem graph. Due to its structural simplicity and computational hardness, it is a very well-studied problem in the field of quantum optimization \cite{Farhi2012Performance, 1411.4028, Zhou2020PRX}.

\paragraph{Handling arbitrary edge weights}

All of the methods discussed below can be applied in the case of unweighted Max-Cut, where all of the edge weights are equal to $1$. However, in order to account for arbitrary edge weights $w_{ij}$, it is necessary to rescale the unweighted graph $\gammav$ parameters as follows~\cite{parameter2023shaydulin}:
\begin{align}
    \boldsymbol{\gamma} \rightarrow \frac{\boldsymbol{\gamma}}{\sqrt{|E|^{-1}\sum_{ij} w_{ij}^2}}.
\end{align}
Such a rescaling applies to all of the methods discussed further in this section when a weighted Max-Cut problem is being considered.

\paragraph{Dweight Method}

The \texttt{Dweight} method builds on the fixed-angle conjecture \cite{wurtz2021fixedangleconjectureqaoa} and uses optimal parameters obtained for $d$-regular tree graphs. We pre-calculated the optimal parameters $\gammav^{\text{tree}}[d]$ and $\betav^{\text{tree}}[d]$ for $1\le p\le6$ and $d\in [1,10] \cup \{20\}$ by optimizing the energy formula $\nu_p(\gammav,\betav,d)$ in \cite{basso_et_al:LIPIcs.TQC.2022.7}. For unseen degrees, we predict the optimal parameters using the following observed scaling of $\gammav^{\text{tree}}[d]$ and $\betav^{\text{tree}}[d]$:
\begin{gather}
\gamma_1^{\text{tree}}[d]  \sim \tan^{-1}\left(\frac{1}{\sqrt{d - 1}}\right), \qquad \gamma_p^{\text{tree}}[d]  \sim \frac{1}{\sqrt{d - 1}} \text{ for } p \ge 2, \nonumber \\
\text{and} \qquad \beta_p^{\text{tree}}[d] - \beta_p^{\text{tree}}[\infty] \sim \frac{1}{d}.
\label{eq:param-scaling}
\end{gather}
Finally, the predicted parameters for any given graph are weighted averages based on the degree distribution:
\begin{equation}
\left(\boldsymbol{\gamma}, \boldsymbol{\beta}\right) = \sum_d w_d \cdot \left(\boldsymbol{\gamma}^{\text{tree}}[d], \boldsymbol{\beta}^{\text{tree}}[d]\right),
\end{equation}
where $w_d$ are weights derived from the graph's degree distribution:
\[ w_d = \frac{d|\{v : \deg(v) = d\}|}{\sum_d d |\{v : \deg(v) = d\}| }. \]
This approach is particularly effective for graphs with heterogeneous degree distributions.

\paragraph{SKatan Method}

The \texttt{SKatan} method is based the fact that $\lim_{d\to\infty}(\sqrt{d}\gammav^{\text{tree}}[d],\betav^{\text{tree}}[d]) = (\gammav^{\SK}, \betav^{\SK})$, where $(\gammav^{\SK}, \betav^{\SK})$ are optimal parameters for the Sherrington-Kirkpatrick (SK) spin glass model~\cite{basso_et_al:LIPIcs.TQC.2022.7,farhi19}.
Following the proposal of parameter setting in \cite{parameter2023shaydulin, Sureshbabu2024parametersettingin}, we rescale the analytic SK optimal parameters using the observed scaling~\eqref{eq:param-scaling} of parameters observed in $d$-regular graphs and information about the average degree $\braket{d}$ of the given graph to predict:
\begin{align}
    \left(\boldsymbol{\gamma}, \boldsymbol{\beta} \right) = \left(\boldsymbol{\gamma}^{\SK} \tan^{-1} \left(\frac{1}{\sqrt{\braket{d} - 1}}\right), \boldsymbol{\beta}^{\SK} \right).
\end{align}

\paragraph{Balanced Approach}
The balanced approach combines elements of \texttt{Dweight} and \texttt{SKatan} to optimize for both regular and irregular graphs. Parameters are computed as a weighted average of the Dweight and SKatan predictions:
\begin{equation}\label{eq:balanced_method}
\left(\boldsymbol{\gamma}, \boldsymbol{\beta} \right) = \alpha \cdot \left(\boldsymbol{\gamma}^{\text{SKatan}}, \boldsymbol{\beta}^{\text{SKatan}} \right) + (1 - \alpha) \cdot \left(\boldsymbol{\gamma}^{\text{Dweight}}, \boldsymbol{\beta}^{\text{Dweight}} \right),
\end{equation}
where $\alpha$ is a tuning parameter that balances the contributions of the two methods, generally set to $0.5$ throughout the rest of this paper.

\begin{figure}[t]
    \centering
    \includegraphics[width=\textwidth]{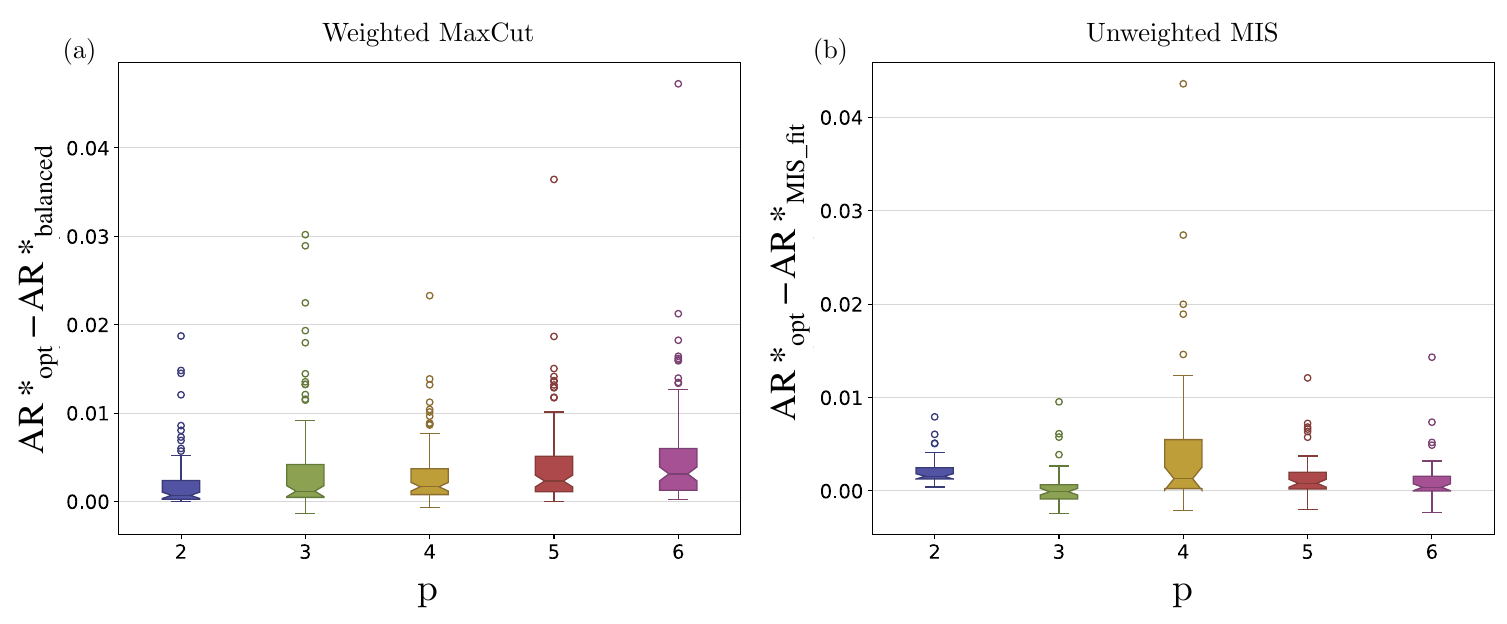}
    \caption{Boxplots evaluating the performance of the `balanced' approach from Eq.~\eqref{eq:balanced_method} for (a) Max-Cut graphs and (b) the MIS fitting prediction method described in Sec.~\ref{sec:angle_prediction_MIS}. We plot the difference in AR* (Eq.~\eqref{eq:approx_ratio_rescaled}) between the states obtained using optimized QAOA angles ($\text{AR*}_{\rm opt}$) and those obtained using either prediction method ($\text{AR*}_{\rm balanced}$ for the `balanced' approach and $\text{AR*}_{\rm MIS\_fit}$ for the MIS fitting approach) across 100 Erd\H{o}s-R\'enyi graphs with $N \in [20, 25]$ nodes and edge probabilities chosen between $0.1$ and $0.9$. In the case of Max-Cut, the edge weights are randomly uniformly generated from the range $[-1,1]$, while the MIS case is unweighted, as described in the text. }
    \label{fig:prediction_method_eval_mis_maxcut}
\end{figure}

We evaluate the `balanced' method in Fig.~\ref{fig:prediction_method_eval_mis_maxcut}, where we compare the rescaled approximation ratio metric from Eq.~\eqref{eq:approx_ratio_rescaled} of states obtained using predicted parameters and those which have been optimized using a classical optimizer. We find that the median error in AR* is  $0.0016$ across all numbers of layers $p$, with some graphs performing even better with the predicted parameters than those which have been optimized. The latter is likely a consequence of the highly non-convex optimization landscape of the QAOA energy.

\subsection{Maximum Independent Set}\label{sec:angle_prediction_MIS}

\begin{figure}
    \centering
    \includegraphics[width=\textwidth]{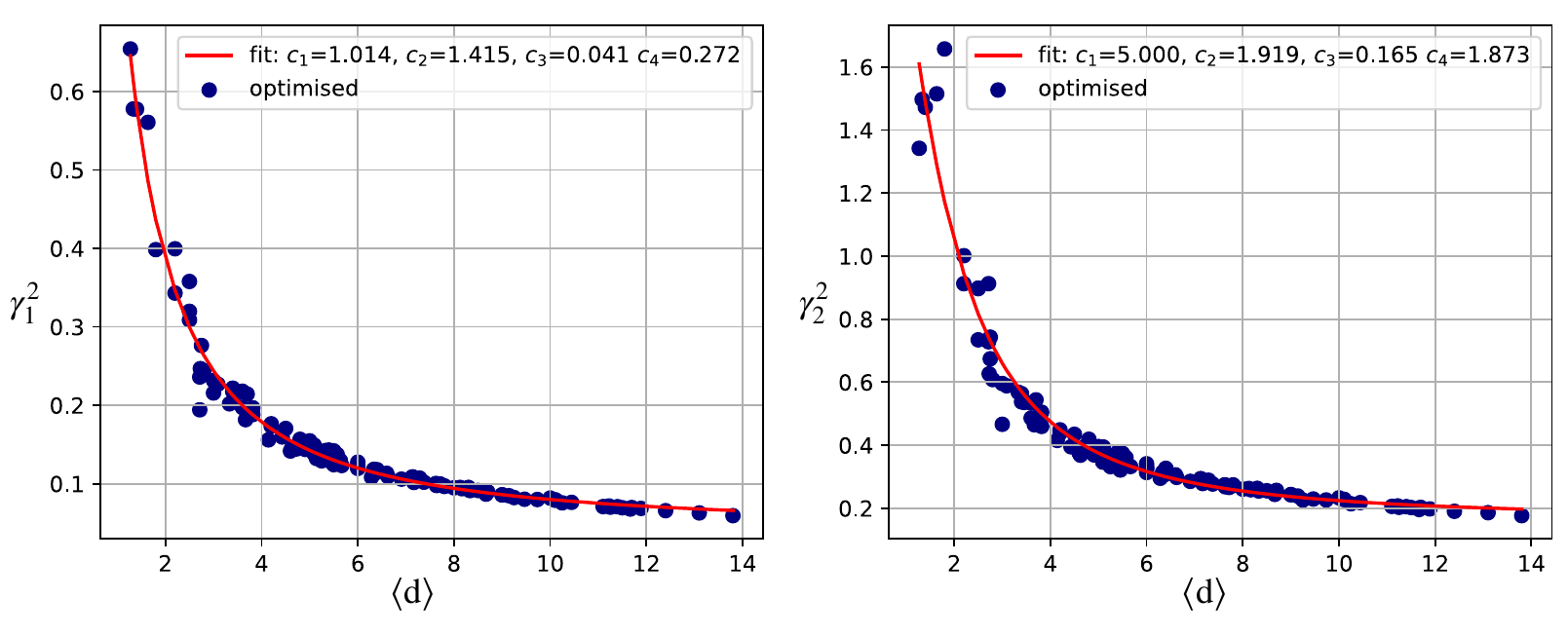}
    \caption{Scatter plots of the optimal and guessed values of $\gamma_j^p$ for total number of layers $p = 2$ versus the average degree of the graph instance $\mean{d}$. Each point corresponds to a different graph, with the navy scatter plot showing the optimized values obtained via an emulation and the L-BFGS-B algorithm as discussed in the text while the red fit shows the prediction model to be used for new instances, as described in Sec.~\ref{sec:angle_prediction_MIS}. The graph instances used are Erd\H{o}s-R\'enyi with $N \in [8, 20]$ nodes and edge probabilities in the range $0.2-0.7$. They are new graph instances, not used for fitting the prediction method.}
    \label{fig:mis_angle_prediction}
\end{figure}

In the case of Maximum Independent Set (MIS), the cost Hamiltonian in Eq.~\eqref{eq:cost_hamiltonian} contains local $Z$ operators on each of the qubits in addition to the $ZZ$ edge interactions. Unlike Max-Cut, it is a model that has been studied far less in the literature and thus lends itself less readily to analysis based on previous results. However, initial investigation appears to indicate that the QAOA angle schedules which minimise the energy of its Hamiltonian from Eq.~\eqref{eq:mis_hamiltonian} contain some structure across different problem instances relating to their average degree. This is exemplified in Fig.~\ref{fig:mis_angle_prediction}, where for $p = 2$ layers of QAOA, the relationship between the average degree of the problem graph and the value of $\gamma$ is either monotonically increasing or decreasing overall in a smooth fashion. 

The case of the plots in Fig.~\ref{fig:mis_angle_prediction} is consistent across the first few layers of QAOA for $p \leq 6$. This allows one to model the functional relationship between the average degree of a graph instance $\langle d \rangle$ and the value of $\gamma_j^p$, where $j = 1,\dots,p$ is the index of the layer and $p$ is the total number of layers in the current instantiation of QAOA. We find that the best fit for each layer $j$ for a $p$-layer QAOA protocol can be modeled via four parameters $\{c_i(j,p)\}_{i = 1,...,4}$:
\begin{align}\label{eq:gamma_fit}
    \gamma_j^p = c_1(j,p) + \frac{c_2 (j,p)}{\mean{d}^{c_3(j,p)} + c_4(j,p)},
\end{align}
where the values of $c_i$ can be determined using a simple fitting algorithm like \texttt{scipy}'s \texttt{curve\_fit}. The resulting set of parameters $c_i$ is used to predict new values of $\gamma$ for previously unseen problem instances. We perform the parameter fitting on a set of 105 Erd\H{o}s-R\'enyi graphs ranging in size from $10$ to $20$ nodes and with edge probabilities set to $0.2-0.7$ for each size.

Such a simple relationship between the optimal values of $\gamma$ and graph degree does not, however, hold throughout all layers for each $p$. As $p$ increases, the first few layers do not deviate much from the functional form in Eq.~\eqref{eq:gamma_fit}, but the final layers lose this clean relationship with $\langle d \rangle$. We treat these extraneous cases by instead using a symbolic regression tool, \texttt{PySR} \cite{cranmerInterpretableMachineLearning2023} to find the best functional fit via a mean-square error loss function. We find that while such a functional relationship is not always possible with a low loss value, the error in angle predictions does not appear to lead to a significant error in the final state energy of the QAOA output.
\newpage
In the case of the mixing Hamiltonian parameters $\beta$, we observe less of an obvious relationship between the degree of the graph instance and their optimal values. However, we note that the variation in $\beta$ across instances is a lot less pronounced than in the case of $\gamma$. Owing to this, for new graph instances, we simply use the average of the optimal $\beta$ obtained for the rounded average degree of the graph:
\begin{align}
    \boldsymbol{\beta}_{\rm new} = \langle \beta_{\rm opt} \rangle_d.
\end{align}
For a new graph with some average degree $d$, we round $d$ to its nearest integer and use the average $\beta_{\rm opt}$ obtained for that degree as a prediction. The maximum degree is capped to $12$ and any graph with higher average degree simply uses the $\beta_{\rm opt}$ value for $d = 12$. This is similar to our approaches in the Max-Cut case, where we assume that for graphs with large degree and with $p < 7$ QAOA layers, the optimal $\beta$ parameters will converge enough for the error to be negligible. 

We evaluate the MIS prediction method in Fig.~\ref{fig:prediction_method_eval_mis_maxcut} on larger graph instances than those used for fitting the prediction functions, as in the case of Max-Cut, and find that it performs well, with a median error in AR* of $0.0008$ across all $p$, occasionally outperforming the classical optimization when initialized from random or linear ans\"atze, as in the Max-Cut case. This is a testament to how difficult it can be to perform variational optimization of QAOA parameters even in the case where a circuit can be simulated classically.  

\subsection{General QUBOs}\label{sec:gnn_prediction}

Most of the results in this work focus on weighted Max-Cut and MIS problem instances -- however, it is fruitful to develop a parameter prediction method for more general QUBOs, where the $Q$ and $c$ in Eq.~\eqref{eq:qubo} can be arbitrarily chosen. This would allow our quantum-enhanced optimization methods to be implemented efficiently for a far larger family of optimization problems, as well as offering insights into the performance of QAOA in new regimes which have not previously been studied.

While there are guiding principles that might help with such a task like in the examples of Max-Cut and MIS -- e.g.\~the decrease of angle magnitude with increasing system size or the relationship between optimal angles and the degree distribution of the problem graph -- the sheer size of the parameter landscape is too large to attempt a similar analysis for arbitrary QUBOs. However, this does not mean that some regression-based or statistical approaches would not be useful, given their effectiveness for the case of Max-Cut and MIS. 

With this in mind, we explore the possibility of using graph-neural-networks (GNNs) \cite{bronstein2021geometric} to predict near-optimal QAOA parameters for arbitrary QUBO problems. GNNs are a family of neural networks that can operate naturally on graph-structured data, reducing the size of the learned function landscape by taking advantage of symmetries in the domain of the training data, which in this case are graphs representing QUBOs. As the optimal QAOA parameters should remain unchanged under a relabeling of the graph nodes, GNNs offer an efficient way to learn the underlying patterns relating QUBO instances to optimal QAOA parameters, with the potential of scaling well across increasing problem sizes. We note that GNNs have been applied to study parameter prediction for QAOA in previous works \cite{deshpande_capturing_2022, liang_invited_2024, xu2025qaoa}, however these studies were limited in their scope both with respect to problem size and the problem types that were explored.

\begin{figure}
    \centering
    \includegraphics[width=0.5\textwidth]{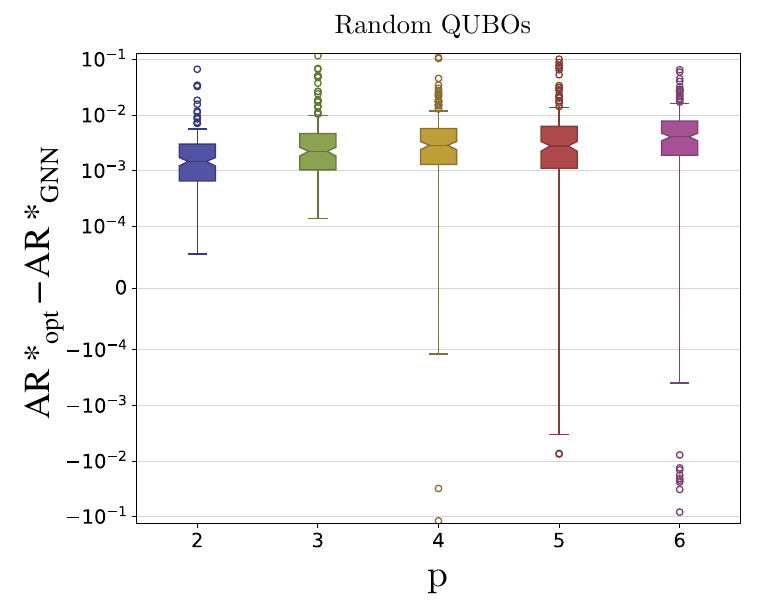}
    \caption{Evaluating performance of GNN predicted parameters against those optimized using a classical optimizer. We train a separate GNN model for each $p$ using pre-optimized parameters for graphs with $N \in [10, 20]$ nodes and use it to predict QAOA parameters for unseen, larger graphs with $N \in [21, 25]$.We compare the energies of the states obtained when using GNN-predicted parameters versus those optimized classically via the boxplots showing the difference between AR* (Eq.~\eqref{eq:approx_ratio_rescaled}) values obtained using either optimized parameters ($\rm{AR*}_{\rm opt}$) or GNN predictions ($\rm{AR*}_{\rm GNN}$) for all QAOA layers across 250 randomly generated QUBO instances, as described in the main text.}
    \label{fig:gnn_evaluation}
\end{figure}

In our implementation, we train a separate GNN for each number of QAOA layers, $p = 2$ up to $p = 6$. We generate training data by optimizing QAOA parameters for Erd\H{o}s-R\'enyi graphs with edge probabilities sampled from the range $[0.1, 0.9]$ for graphs with up to $20$ nodes. The edge and node weights are randomly generated based on a set of structured and unstructured QUBO problems, including Max-Cut and MIS, with weight amplitudes constrained to be in $[-1,1]$. We find that the GNNs produce high-quality predicted QAOA parameters even when trained on only several thousand graphs for each model, scaling well on test graphs which are larger than those in the training data. 

In Fig~\ref{fig:gnn_evaluation} we plot the difference in AR* between states obtained when using optimized QAOA parameters ($\rm AR^*_{\rm opt}$) and those predicted by the GNN models ($\rm AR^*_{\rm GNN}$) across 250 random Erd\H{o}s-R\'enyi QUBO graphs with $N \in [20,25]$ nodes and edge probabilities sampled from $[0.1, 0.9]$. The QUBO node and edge weights making up $Q$ and $c$, all in the range $[-1,1]$ are generated by either sampling all of them from a uniform distribution in the total range, or by fixing either the node/edge weight while generating the edge/node weights randomly, respectively. While these graphs do not capture the full range of possible QUBO problems, they lend some credence to the ability of our GNN models to predict near-optimal parameters in a wider range of QUBO structures. The median error in AR* across all graphs in the plot is $0.0014$. More details on this approach can be found in Appendix~\ref{sec:appendix_predicting_gnns}.

\section{Qubit mapping and routing}
\label{sec:hardware}

When QAOA is applied to a graph problem like Max-Cut or MIS, each qubit in the quantum computer corresponds to a vertex in the graph $G=(V,E,w)$. Thus, each QAOA cost Hamiltonian layer consists of 2-qubit quantum gates $e^{i w_{jk} \gamma^i Z_j Z_k}$ for each edge $(j,k) \in E$, where $w_{jk}$ is the weight associated with the edge in a weighted graph problem (with $w_{jk} = 1$ in the unweighted case) and $\gamma^i$ the $i^{\rm th}$ variational parameter corresponding to the $i^{\rm th}$ layer of QAOA. In the case of MIS, the cost Hamiltonian layer will have an additional set of local $e^{i \gamma^j Z_j}$ operators corresponding to each vertex of the graph $j \in V$. We will call each of the 2-qubit quantum gates a \emph{term}. As the terms for cost Hamiltonians with the type of structure found in Eq.~\eqref{eq:cost_hamiltonian} commute, they can be implemented in any order. However, most quantum hardware platforms impose connectivity constraints such that not all edge terms can be implemented without the use of swap operations to re-organise the terms (or, equivalently, the qubits). The order in which the terms are implemented and the swap operations required is known as the \emph{routing} problem.

Furthermore, we also need to assign each vertex in the problem instance to a hardware qubit on the device. Let the connectivity of the hardware (across which qubits gates can be implemented) be specified by a graph called the hardware graph. The \emph{qubit mapping} problem of how best to associate vertices with qubits can be solved in a number of ways, and the routing algorithms below are agnostic to which method is used.

\subsection{Qubit Mapping}

One effective and straightforward solution is to use an algorithm for the routing problem itself to generate a qubit mapping~\cite{li19}. Starting from any initial qubit mapping, the final mapping of vertices to qubits at the end of the circuit produced by the routing algorithm can be used as the initial qubit mapping for another iteration of the routing algorithm. What remains is to devise an initial mapping which we can pass to the routing algorithm. The simplest method one can use to do this is to generate a random assignment of vertices to qubits. A more efficient approach may be to provide the routing algorithm with a more informed initial guess which may require less effort to optimize. The next sections discuss methods to do this.

\subsubsection{Line Laplacian}

The first approach aims to assign the vertices of the problem graph to the hardware graph in a way that minimises the distance between any two qubits which need to interact~\cite{lin21}.
Let $G(V, E)$ be an undirected graph, then the Laplacian matrix $L \in \mathcal{M}_{|V| \times |V|}(\mathbb{R})$ is defined to be: 
\begin{align}
    L_{i,j} = \begin{cases}
        \deg(v_i) & \text{if } i = j \\
        -1 & \text{if } (v_i, v_j) \in E \\
        0 & \text{otherwise}
    \end{cases}.
\end{align}
This is a symmetric positive definite matrix, with smallest eigenvalue $0$ with eigenvector $\mathbf{1}$ the all-ones vector. Assuming the graph is connected the 0-eigenspace is 1 dimensional, and the unit eigenvector with the second smallest eigenvalue is known as the \emph{Fiedler vector}, which can be used for partitioning and clustering of graphs \cite{fielder_graph_connectivity}.

Suppose we want to map the vertices of the graph $G$ to a line, which can be represented as a function $f: V \to \mathbb{R}$, such that we minimise the distance between connected nodes. This is a minimisation problem as described in \cite{spielman_graph_embedding}:
\begin{align}
\begin{split}
    \arg \min_{f} & \  (f(v_i) - f(v_j))^2 \\
    s.t. & \sum_{v \in V} (f(v))^2 = 1, \sum_{v \in V} f(v) = 0.
\end{split}
\end{align}
The first constraint normalises the output so the resulting positions on the line are not all 0, and the second constraint means that the resulting positions are not all equal. Writing $\mathbf{f}$ for the vector $(f(v_1), ..., f(v_n))$, we can rewrite the objective function as $\mathbf{f}^T L \mathbf{f}$, and so the solution to this optimization problem is precisely the Fiedler vector.
We can use this to determine a mapping from the problem graph $G$ of Max-Cut to the hardware graph $H$. The process proceeds as follows:

\begin{enumerate}
    \item Find a sequence $S = (s_1, ..., s_{|V_G|})$ of distinct vertices in $H$ of length $|V_G|$ such that any adjacent pair of vertices in the sequence are connected by an edge in $H$.
    \item Determine the Fiedler vector $\mathbf{f}$ of the graph $G$, treated as an unweighted graph.
    \item Order the vertices of $G$ by the corresponding component of $\mathbf{f}$.
    \item Assign the $k^{th}$ vertex in the ordered sequence to $s_k$.
\end{enumerate}

This will place the graph vertices on a line, such that the sum of the distances between any two qubits which need to interact is minimised. This works well for hardware graphs which are sparsely connected, as it does not take into account the full connectivity of the graph $H$.

\subsubsection{Quadratic Assignment Problem}

An alternative approach, also implemented in 2QAN~\cite{lao22}, is to try to assign vertices in the problem graph to qubits in a way that maximises the total number of edges which overlap. Let $V_G$ and $V_H$ be the vertices in the problem graph and hardware graph respectively, and $d_G: V_G \times V_G \to \{0, 1\}$ and $d_H: V_H \times V_H \to \{0, 1\}$ be the indicator functions of $E_G$ and $E_H$ respectfully. We want to find the maximum value over all injections $f: V_G \to V_H$ of:

\begin{align}
    \sum_{u, v \in V_G} d_G(u, v) d_H(f(u), f(v)).
\end{align}

This is an instance of the \emph{quadratic assignment problem} \cite{umeyama_qap}, a combinatorial optimization problem which aims to assign facilities to locations in a way that minimise the cost of flowing between facilities. For the precise reduction, we require $|V_G| = |V_H|$ and a bijection between the vertex sets, which is not always the case in our context, but we can expand $G$ by adding extra unconnected nodes that can map to anywhere on the hardware. This is still an NP-hard problem~\cite{papadimitriou_qap_np_hard}, but there are heuristic methods that rapidly obtain an approximately optimal solution, for example in \cite{vogelstein_quadratic_assignment}. Also as we are using this as an input to the routing algorithm, we can tolerate a suboptimal solution and improve it with the routing method.

\subsection{Routing}

Once we have the initial layout of the vertices on the hardware graph, we need to apply swap operations to reorganise the qubits so that the required interactions are possible on the constrained hardware. We provide two methods to do this: 

\subsubsection{Greedy algorithm}

Here, we describe an approach based on a modification of an idea first presented in~\cite{clinton24} for implementing the terms in a fermionic quantum Hamiltonian using fermionic swap networks.

Once a mapping is chosen, our procedure generates a quantum circuit which is made up of a sequence of swap gates, interspersed with terms. At any given time, the procedure maintains a list of terms that have not yet been implemented. Whenever terms can be implemented, because the corresponding qubits are connected to each other in the hardware graph, we immediately implement the terms and remove them from the list of terms.

Let the distance $d(T)$ of a term $T$ be defined as the shortest path in the hardware interaction graph between the two qubits on which it acts. If the distance is 1, then the term can be implemented immediately; otherwise, at least one swap is needed to implement it. Then we define an overall distance function $D$ as
\[ D = \sum_{T} d(T)^q, \]
for some real number $q$ which is a parameter of the algorithm. In practice, taking $q=1$ -- in which case $D$ is just the total distance of terms -- is already quite effective. However, choosing $q<1$ can be advantageous by prioritising bringing close terms together rather than those at a greater distance.

The algorithm proceeds as follows:

\begin{enumerate}
\item Implement all terms at distance 1. If there are no terms remaining, terminate the algorithm.
\item For all possible swaps (corresponding to edges in the hardware graph) compute the overall distance $D$ following applying that swap to all of the terms.
\item If there exists a swap which reduces the overall distance, apply the swap that minimises the overall distance and go to step 1.
\item If there is no such swap, apply a swap that reduces the distance of the term that was the lowest distance, and go to step 1.
\end{enumerate}

The final step of the algorithm is a fallback in case the best swap leaves the total distance unchanged (for example, by decreasing the distance of one term, but increasing the distance of another).

The above algorithm generates a quantum circuit which is expressed as a sequence of terms interspersed with swap operations. On typical quantum hardware, swap operations are implemented via a sequence of three controlled-not (CNOT) gates. In cases where a swap operation occurs before or after a ZZ term on the same qubits, advantageously the combination of the swap and the term can also be implemented via three CNOT gates. This is discussed in more detail in Appendix~\ref{sec:graphs_zz_swaps}.

A related algorithm and software tool, termed 2QAN, for implementing terms in a Hamiltonian in an arbitrary order using swap operations was presented in~\cite{lao22}; this algorithm also uses the above optimization technique for merging swap operations with terms. However, our approach differs by using an overall distance function which takes into account the distances of all terms, whereas the algorithm of~\cite{lao22} always aims to reduce the distance of the lowest-distance term (which in our algorithm is just a fallback). Thus our approach may be expected to enable large numbers of terms to be implemented more quickly.

\subsubsection{A-star}

An alternative approach is to formulate this routing problem as a graph traversal problem, and apply search methods such as A* (A-star)~\cite{hart_a_star} to solve it. These heuristic methods are very general, and have been used in the optimization of quantum circuits before for the qubit mapping problem \cite{zulehner_a_star_qubit_mapping}. For a more detailed description of the A* method, see Appendix \ref{sec:a_star_description}. We begin the search from the start node $s$, which is the empty circuit with the logical vertices in $G$ being assigned physical qubits in $H$, and we search through a graph containing partial circuits containing a subset of the interactions we want to apply and swaps to make them possible. A partial circuit is in the set of target nodes $T$ if every edge in $G$ corresponds to an interaction in the circuit which is valid with respect to the connectivity of $H$. Equivalently, if we attach a list of unimplemented interactions to each partial circuit in $V$, we have reached a target node when this list is empty. The algorithm will search through partial circuits until a node in the target set $T$ is reached with low path cost.

\pagebreak

Given a problem graph $G$ and hardware graph $H$, the directed weighted graph $S(V, E, W)$ we are exploring is defined as:
\begin{itemize}
    \item The vertex set $V$ is the set of partial circuits containing a sequence of swap gates interspersed with terms, such that all the swaps are valid edges in $H$ and all of the interactions correspond to logical edges in $G$.
    \item There is an edge $(C_1, C_2)$ between two partial circuits $C_1$ and $C_2$ if we can form $C_2$ from $C_1$ with a set of disjoint swaps in $H$, followed by some interactions from $G$.
\end{itemize}
This is an exponentially sized graph as there are many possible partial circuits and so we are unable to store it all at once, but we are able to efficiently compute the neighbours of a given node. A nice feature of $S$ in this problem is that it is a tree, as for every partial circuit there is only one path to reach it - the sequence of swap and interaction layers it contains. This means we do not have to worry about revisiting nodes during our search.

At any given point in the search, the A* algorithm will pick the next node to explore based on the cost function $g: S \to \mathbb{R}$ and heuristic $h: S \to \mathbb{R}$, which is an estimate for the cost required to reach a target node in $T$. There are multiple cost functions and heuristics that we can use, which have effects on the optimality and runtime of the algorithm. The choice of cost function is also dependent on what the optimization target is, whether it is gate count or depth of the circuit. This means that the A* algorithm encompasses many different graph search methods by choosing different functions for $g$ and $h$. 

For the cost $g$ of a partial circuit, we can simply use the total number of swaps required in the circuit to run the quantum algorithm. This is a good metric when the number of gates in the circuit is important, as every swap will require 3 CNOT gates to implement on typical quantum hardware. Alternatively, we can group disjoint swaps together that can act at the same time to get an estimate of the depth of the partial circuit. Furthermore, we can perform the same optimization of merging swap operations and terms which 2QAN and the greedy method make use of.

The choice of heuristic $h$ is a difficult question that comes with many trade-offs. The heuristic that we used is the same total distance measure used in the greedy approach:
\[ D = \sum_{T} d(T)^q, \]
for some real number $q$ which is a parameter of the algorithm, most of the time we set $q=1$. This is not an admissible heuristic, as one swap can make multiple interactions closer together in the hardware, and therefore $D$ may potentially overestimate the number of swaps required. We expect further improvements in both runtime and optimality can be made in this framework by choosing better heuristics.

\section{Error Mitigation}\label{sec:error_mitigation}

Due to the noisy nature of near-term quantum devices, the samples obtained when running QAOA are likely to suffer from errors, leading to suboptimal, higher-energy outputs. We address this issue with several error-mitigation techniques which use both information about the physical error rates of the quantum device as well as expected properties of high-quality solutions. Many error-mitigation methods are tailored towards improving estimates of the expectation value of some observable. For our purposes, we require error-mitigation techniques which can instead improve the measured samples directly, making many popular methods like Zero-Noise-Extrapolation~\cite{9259940} or Clifford-Data Regression~\cite{czarnik_error_2021} less applicable. However, other techniques, like readout error mitigation, can still be implemented to improve the quality of the samples. We note here that due to the sample overhead, the recent method of~\cite{liu2025quantum} for sampling error mitigation is less applicable to our scheme.

\subsection{Readout error mitigation}\label{sec:readout_error_mitigation}

We implement a well-known technique for readout-noise mitigation which involves applying the inverse of a noise calibration matrix to the output sample distribution. This noise calibration matrix is constructed using an efficient protocol which requires an estimate of readout error in the computational basis of each qubit involved in the quantum circuit \cite{maciejewski_mitigation_2020}. More precisely, for each qubit one needs to construct a `correction matrix':
\[ \Lambda = \left[\begin{array}{cc}
    1-p & q \\
     p & 1 - q
\end{array}\right]\]
where $p,q \in [0,1]$ are the probabilities of incorrectly measuring a $\ket{0}$ and $\ket{1}$ state respectively. These probabilities can be obtained by simply running two simple circuits on the qubits used in the computation and collecting their statistics: a circuit initialised to the all-$0$ state and one initialised to an all-$1$ state via the application of $X$ operators on each of the qubits. One may also obtain $p$ and $q$ from calibration data of the quantum device, if a provider of said device allows access to the information and performs such calibrations on a regular basis.

The inverse of this matrix, $\Lambda^{-1}$, tensored together for all qubits, can be used to correct for readout errors, as its action on the output measurement distribution of a quantum circuit inverts the classical bit-flip errors estimated by $\Lambda$. In practice, however, this method is quite computationally intensive as the tensor product of each $\Lambda^{-1}$ for every qubit grows exponentially with the number of qubits. 

In our experiments, we use an efficient implementation of the correction matrix inversion method which relies on the sparsity of the output bit-string distributions of the quantum circuit in order to reduce both memory and runtime requirements, first presented in \cite{yang_efficient_2022} by Yang et al. In their work, the authors introduce several readout-error-mitigation `filters' based on previous innovations like the SGS algorithm~\cite{PhysRevLett.108.070502} and new improvements. In our experiments, we implement the \emph{least norm} filter (LNF) presented in their work, which numerically showed the most favourable trade-off between improvement in output bit-strings and the computational resources required for implementation.

\subsection{QAOA bit-string refinement}\label{sec:em_filters}

While low-energy bit-strings for an NP-hard problem like Max-Cut or MIS are difficult to generate, it is fairly straightforward to compute the energy of a bit-string once it is obtained. This allows us to employ several further `filtering' techniques for the bit-string distributions obtained from running QAOA on hardware. 

\begin{itemize}
    \item \textbf{Energy-filter}: We compute the energies of all the samples obtained from QAOA efficiently and keep only the 10\% of bit-strings with the lowest energy. 
    
    \item \textbf{Frequency-filter}: We keep only the bit-strings which have the highest frequency in the QAOA output distribution, based on some threshold frequency (set to $0.05\%$, or at least 5 occurrences in 10000 samples). If no bit-strings occur with a frequency above the threshold, then the threshold is lowered until some do. If the QAOA output is a uniform distribution over bit-strings, no filtering is applied. This filter is informed by the intuition that a `good' QAOA output should concentrate around bit-strings that either are or approach the solution string. On the other hand, if errors occur on the QAOA output, each resulting bit-string is likely to occur with low probability. We note that this filter is not applicable in cases where the output set of samples is a uniform distribution, either due to high noise levels or a smaller sample size.
    
    \item \textbf{Hamming-filter}: As in energy-filtering, we select only the lowest-energy bit-strings, but a lower percentage ($1\%$) and choose another $9\%$ of strings closest in Hamming distance to those low-energy strings. This filtering is informed by the local search structure of the classical algorithms which we aim to accelerate. Strings which have low Hamming distance to the solution, or at least to a high quality string, are likely to lead to faster convergence, even if they have higher energies.
\end{itemize}

\section{Experimental Results}
\label{sec:results}

We have implemented the above approach in a full software pipeline that allows for finding the solutions of large and complex Max-Cut instances using emulated or real quantum hardware.

\subsection{Emulated results}\label{sec:emulated_results}

\begin{figure}[t]
\centering
    \includegraphics[width=0.82\linewidth]{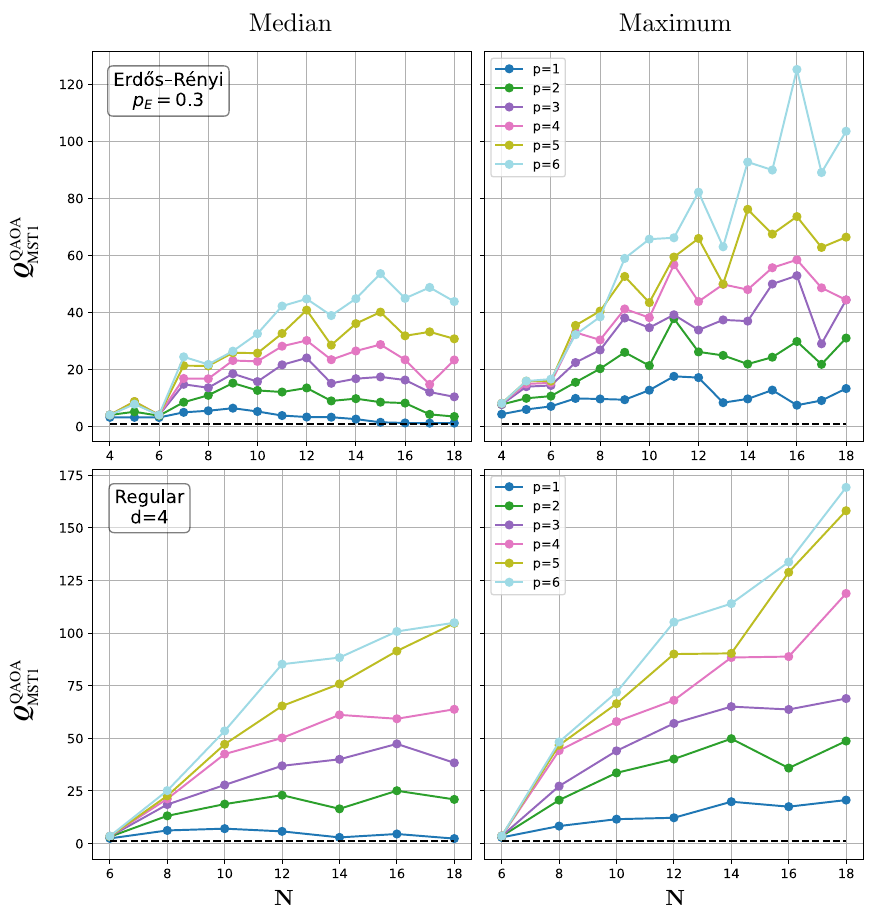}
\caption{Q-factor when warm-starting the PalubeckisMST1 algorithm implemented in MQLib~\cite{dunning_what_2018} for unweighted Max-Cut with samples obtained using classical emulation of QAOA. We investigate Erd\H{o}s-R\'enyi graphs with edge probability $p_E = 0.3$ and random-regular graphs with degree $4$. For each graph type and graph size $N$, we generate $10$ random graphs. We use the `balanced' approach from Sec.~\ref{sec:maxcut_angle_prediction} to predict the QAOA parameters and generate $1000$ samples from the classical emulator. The Q-factor (Eq.~\eqref{eq:qfactor}) is estimated using an iterations cut-off of $T_{\rm total} = 5\times10^{2}$ as this is sufficient to always reach the optimal solution for all problem instances of this size.}
\label{fig:palubeckis_er_0.3_reg_4}
\end{figure}

In order to test our method, we first explore the unweighted Max-Cut problem using a classical emulator of a quantum computer. After sampling bit-strings from the QAOA algorithm, we use them to warm-start the PalubeckisMST1 algorithm implemented in MQLib \cite{dunning_what_2018} and compare its performance to initialising with random samples by using the Q-factor from Eq.~\eqref{eq:qfactor}. We refer to the metric as $Q^{\rm QAOA}_{\rm MST1}$ to denote the fact that we are comparing expected runtimes when warm-starting with QAOA samples versus random samples, as in the case of the bare PalubeckisMST1 algorithm from Table~\ref{tab:mqlib_warmstarts}. In order to estimate $Q^{\rm QAOA}_{\rm MST1}$, we perform Max-Cut optimizations on each graph instance using both random initial bit-strings and those obtained from QAOA. We keep track of the number of iterations required in each case to reach the optimal cut for the graph and used those statistics to estimate the expected runtime of the algorithm for a given set of initial strings.

In Figure \ref{fig:palubeckis_er_0.3_reg_4} we show the results obtained for Erd\H{o}s-R\'enyi graphs with edge probability $0.3$ and random regular graphs with average degree $4$, for graphs with between $4 \leq N \leq 18$ nodes and angles predicted using the `balanced' method from Sec.~\ref{sec:parameters}. We see a positive scaling as we increase both the number of vertices in the problem graph and increase the number of layers of QAOA, achieving a Q-factor of up to 175 for the largest graphs on $18$ vertices. The regular graphs perform better, where the median speedup factor for Erd\H{o}s-R\'enyi graphs is at most 60, compared to over 100 for regular graphs. This may be due to the `balanced' approach for predicting angles being partially based on graphs of constant degree. As the problem size increases, we also see for small numbers of QAOA layers $p$ the Q-factor starting to decrease again, which could be because the ansatz is not expressive enough to provide samples which are useful as a warm start. This is especially visible for the Erd\H{o}s-R\'enyi graphs. For a full set of results, including experiments accelerating the FESTA2002VNSPR algorithm, see Appendix \ref{sec:appendix_further_emulation}.

Overall, these results indicate that, on a perfect quantum computer, we should be able to achieve a significant speedup by warm-starting classical heuristics on a large class of problem instances.

\begin{figure}[t]
\centering
    \includegraphics[width=\linewidth]{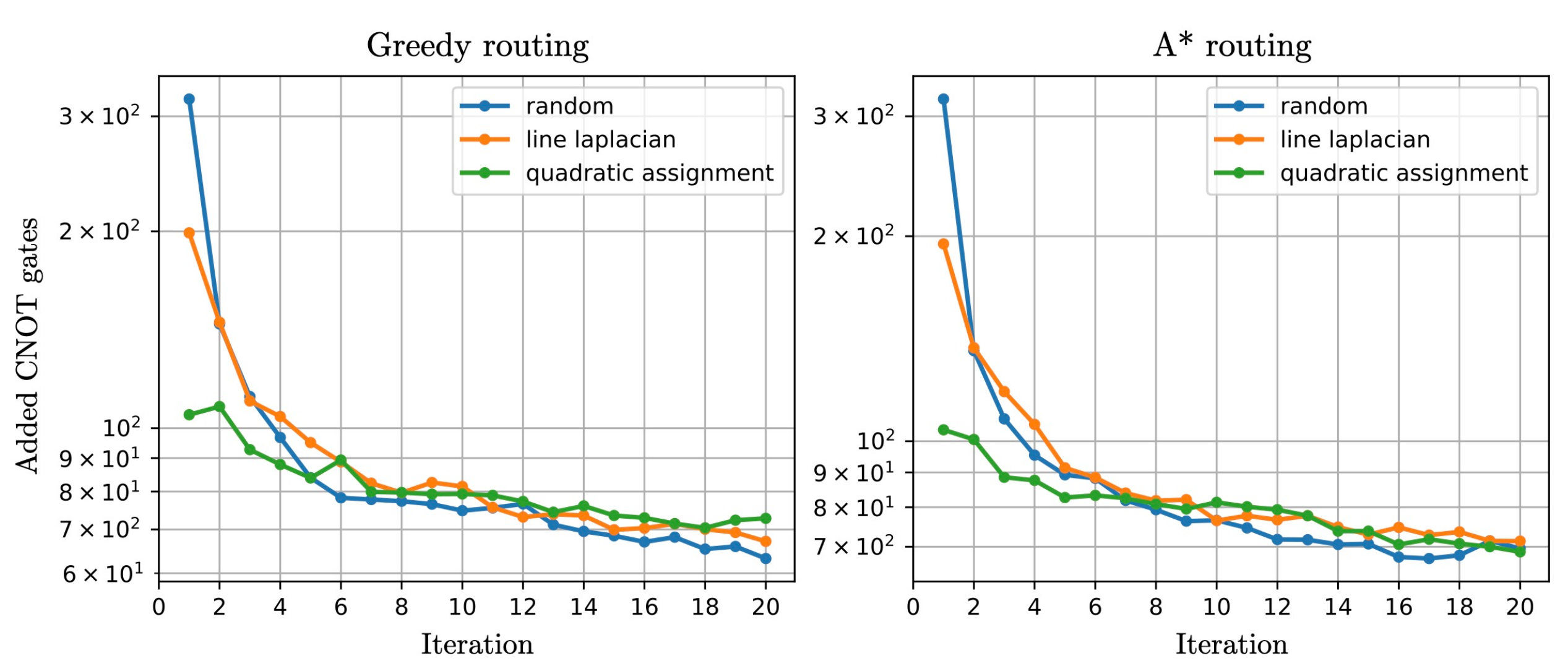}
\caption{The number of additional CNOT gates required to implement one layer of QAOA averaged over random 3-regular graphs with between $20$ and $50$ vertices on quantum hardware with grid connectivity on a $10 \times 10$ grid. Each line is averaged over all graph sizes, with $5$ graphs of each size. With both the greedy (left) and A* heuristic (right) routing algorithms, we see for the first iteration a $38\%$ decrease for the line laplacian layout and $67\%$ decrease for the quadratic assignment layout in the number of extra CNOTs required. To reach the same number of extra CNOTs as the quadratic assignment method after 2 iterations, it takes the random initial layout $4$ iterations for the greedy routing algorithm and $6$ iterations for the A* routing algorithm. After around 8 iterations all three initial layout methods have converged and give similar requirements.}
\label{fig:mapping_comparisons}
\end{figure}

\subsection{Circuit Optimization Results} \label{sec:circuit_optimisation_results}

We first compare the performance of different initial vertex-to-qubit mappings. The task we choose is to lay out random 3-regular graphs with between $20$ and $50$ vertices on a quantum computer with a $10 \times 10$ square lattice connectivity. We run both the greedy and A* heuristic routing methods for $20$ iterations, each time using the final mapping found by the routing algorithm as the initial mapping in the next iteration as described in Sec.~\ref{sec:hardware}. When computing the CNOT cost of implementing the circuit, we also use the optimization of merging swaps and interaction terms. In Fig.~\ref{fig:mapping_comparisons} we see that when the number of iterations is less than $5$, the quadratic assignment mapping method results in the fewest number of extra CNOT gates. After around $5$ iterations, all three methods begin to give similar CNOT counts as the mappings converge. This means that the quadratic assignment method with 2 iterations gives the same number of added CNOTs as a random initial layout with around 6 iterations. We also see that while most of the improvement in CNOT count occurs at the first few iterations, with the greedy routing method marginal gains can still be made even at $20$ iterations, whereas the A* heuristic converges to the optimal mapping by around $15$ iterations.

\begin{figure}[t]
\centering
    \includegraphics[width=\linewidth]{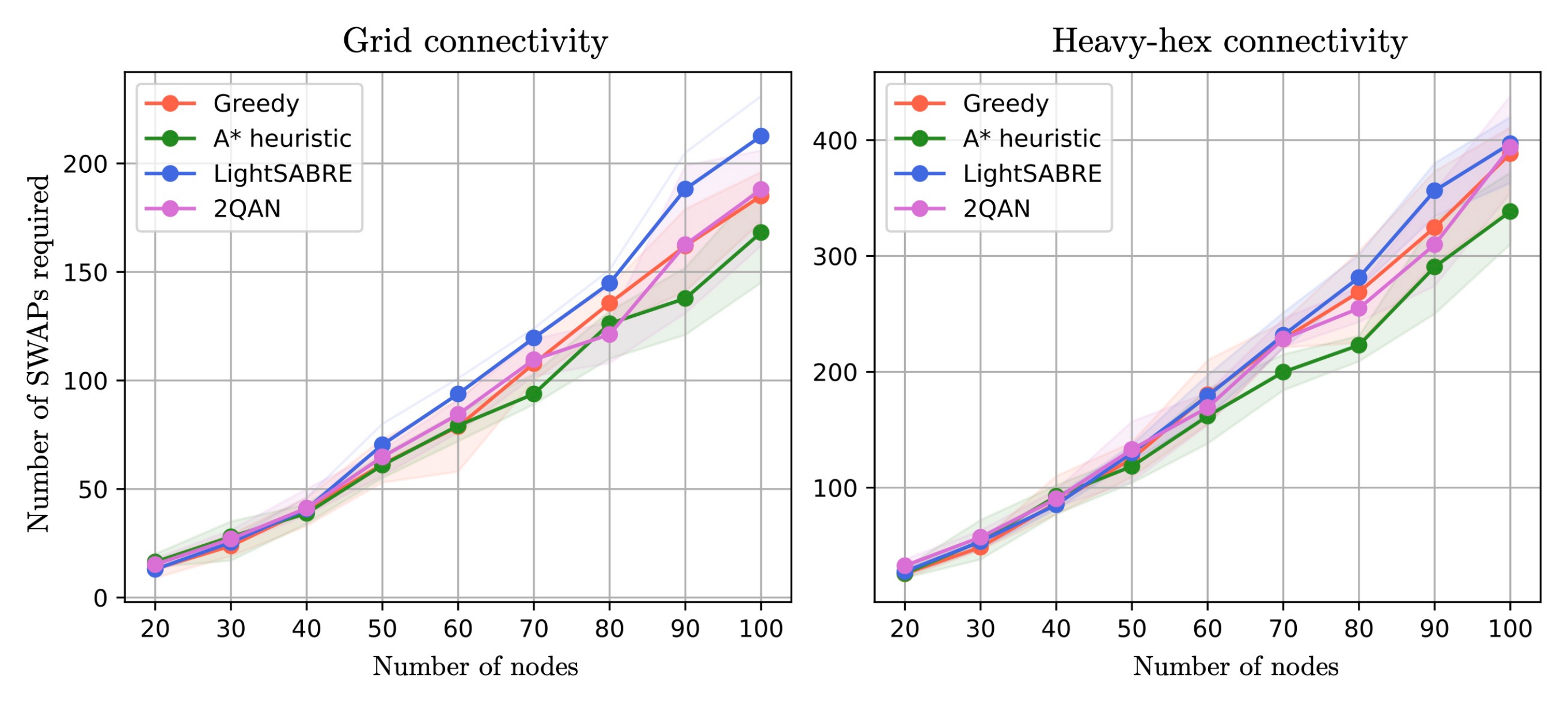}
\caption{The number of swaps required to implement one layer of QAOA for random 3-regular graphs with between $20$ and $100$ vertices, using the greedy algorithm and the A* heuristic using $q=1$ in the distance computation and starting with the quadratic assignment initial vertex to qubit mapping, with comparisons to the existing methods 2QAN~\cite{lao22} and LightSABRE~\cite{zou24_sabre}. Each problem graph was transpiled onto a $(12 \times 12)$ grid connectivity and the 156 qubit IBM Heron heavyhex connectivity. Each point on the line is averaged over 5 graphs, with the shaded area showing the minimum and maximum costs found for a given graph size. The best performing method is the A* heuristic in both connectivities. For graphs with 100 nodes, in the heavyhex connectivity we see a reduction in the number of swaps on average required of 14.7\%, 14.4\% and 12.9\% and for the grid connectivity a reduction of 20.9\%, 10.5\% and 9.1\%, compared to LightSABRE, 2QAN and the greedy algorithm respectively.}
\label{fig:routing_comparisons_3reg}
\end{figure}

\pagebreak

Next we compare the complexity of the quantum circuits obtained using our routing methods. We compare these for the task of implementing one layer of the QAOA algorithm on a quantum computer with both $(12 \times 12)$-grid connectivity and the IBM Heron heavyhex connectivity on 156 qubits, for example the \verb|ibmq_fez| device. We initialise the vertex-to-qubit mapping with the quadratic assignment layout, and iterate the routing procedure 10 times using the final layout as the initial layout for the next iteration. We compared our methods against two existing mapping and routing software tools: 2QAN~\cite{lao22}, developed to implement swap networks where the gates can appear in an arbitrary order, and LightSABRE~\cite{zou24_sabre}, one of the IBM Qiskit transpilation techniques. We generate problem graphs with between $20$ and $100$ nodes, averaging over 5 graphs for each problem size, and computed the number of swaps required to implement one layer of the QAOA circuit for each of the four methods, without using the merging swaps and interactions optimization to make a fair comparison to LightSABRE which does not exploit this. In Fig.~\ref{fig:routing_comparisons_3reg} we see that the A* heuristic method performs the best for problems with $N > 50$ vertices, requiring up to 20\% fewer swaps than the other methods. As expected, LightSABRE performs slightly worse than the other methods, which are all specifically for a QAOA circuit where the terms in each layer can be implemented in any order. There is approximately a 10\% reduction in the number of swaps compared to the greedy method, indicating that the additional complexity of being able to backtrack does provide more optimised circuits for these problems. For more details and a more extensive set of results, see Appendix \ref{sec:appendix_further_routing}.

\subsection{Hardware results}

In this section we present three sets of results on different problem families where the classical heuristics were warm-started by samples obtained from quantum hardware. We begin with the simple case of solving Max-Cut on line graphs, as a benchmark, follow it up with Max-Cut on graphs which are SWAP-enhanced versions of the heavy-hex topology and finally, demonstrate solving Max-Cut and MIS for a wide range of Erd\H{o}s-R\'enyi and random-regular graph instances.

\subsubsection{Max-Cut on line graphs}\label{sec:maxcut_line_graphs}

Line graphs constitute a good benchmark for quantum hardware as they are easily mapped onto most quantum device topologies and require low depth circuits even for large scale problem instances. While the solution to Max-Cut on a line graph can be easy to compute with a greedy approach due to the simple structure of the problem, many classical local search heuristics, like the tabu search of PalubeckisMST1, do not inherently exploit this structure. This allows us to explore the potential speedups provided by quantum warm starts for such algorithms.

\begin{figure}
    \centering
    \includegraphics[width=\textwidth]{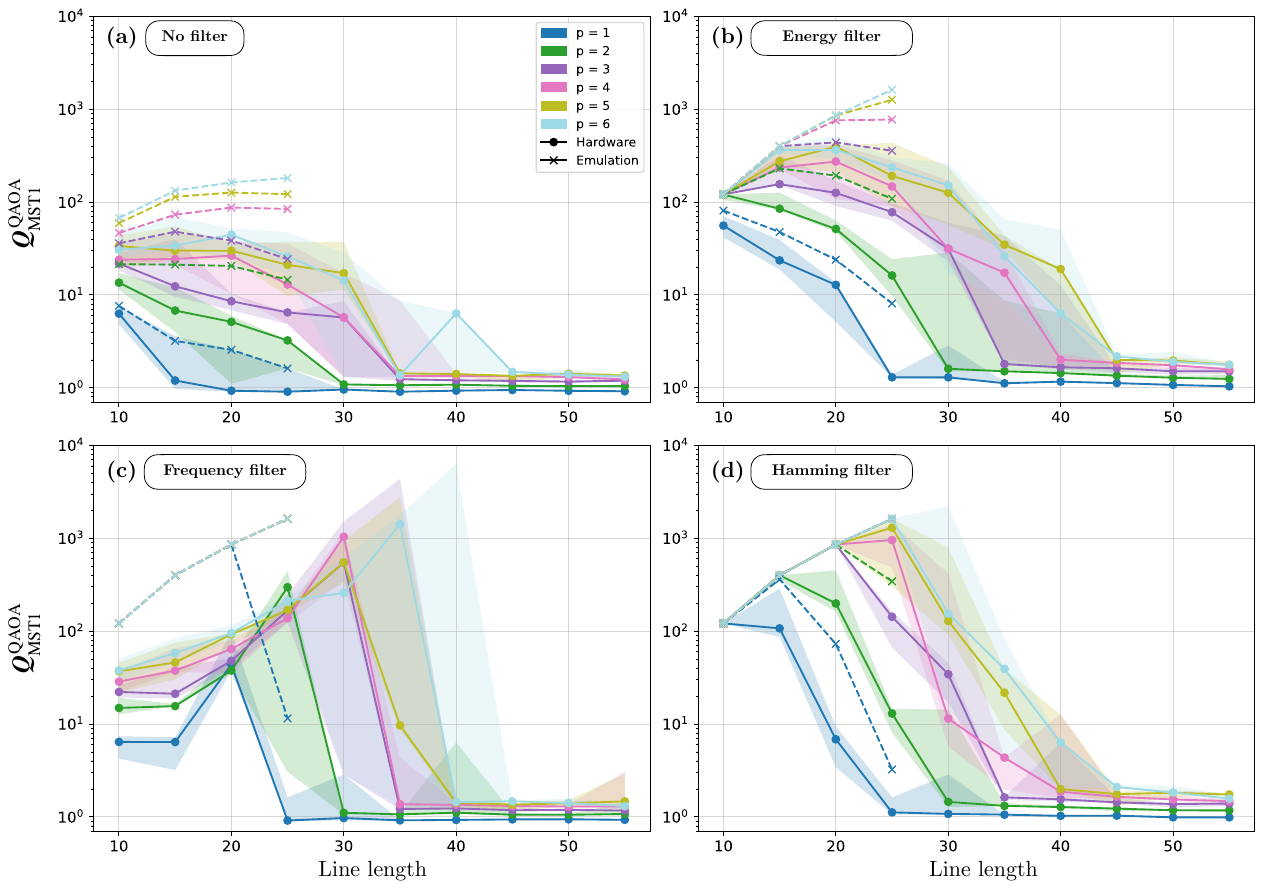}
    \caption{Speed-ups for solving Max-Cut on line graphs with quantum enhanced optimization. We plot the Q-factor for PalubeckisMST1 when started with QAOA samples versus random initialization across different number of circuit layers $p$, where the optimal parameters are predicted using the `balanced' method from Sec.~\ref{sec:parameters}. Each plot shows the median Q-factor with the maximum and minimum speedup obtained shown in the shaded regions, when (a) samples with no filter are used, (b) only the $1000$ lowest energy samples are used, (c) the frequency filter or (d) the Hamming filter is applied to the samples. In each case, readout error mitigation is used on the hardware samples as described in the main text. We plot both the hardware results (solid plot, circles) and samples obtained from emulation (dashed plot, crosses) for comparison.}
    \label{fig:ibm_maxcut_line_graph}
\end{figure}

We investigate the performance of quantum-enhanced optimization in solving unweighted Max-Cut on line graphs ranging from $10$ to $55$ nodes, with QAOA samples obtained from \texttt{ibmq\_torino}, \texttt{ibmq\_fez} and \texttt{ibmq\_marrakesh} devices, as well as comparing to samples obtained in emulation for system sizes up to $25$ nodes. In Fig.~\ref{fig:ibm_maxcut_line_graph} we plot the median, min and max Q-factor (Eq.~\eqref{eq:qfactor}) obtained when running the PalubeckisMST1 tabu search warm-started with QAOA samples, obtained across all three runs on quantum hardware, as well as classical emulation results. In all plots, the hardware samples are error-corrected using the LNF readout error mitigation technique described in Sec.~\ref{sec:error_mitigation}. The Q-factor is obtained across $1000$ optimization runs initialized with either random strings or QAOA samples. 

We find that overall, for $p \geq 2$, the QAOA warm starts always give a Q-factor above $1$, even for the largest graphs and without any additional post-processing. We also investigate the effects of the various post-processing filters on the sample distribution used for warm-starting. We find that the Hamming filter performs best across all $p$, until it is overcome by noise at around $30$ nodes. The frequency filter grows in performance exponentially until it is again dampened at around $35$-$40$ nodes, after reaching a several-thousand-fold speed-up. We can attribute this increase to the fact that the pool of higher-frequency bit-strings gets smaller as the noise in the hardware samples increases and that the frequency filter picks out high quality bit-strings. This smaller pool of high-quality warm starts leads to a higher likelihood of the classical algorithm converging faster to the optimal solution. This effect further explains the shift in speedup improvement from using energy-filtering to frequency-filtering as the problem size increases: the energy filter is consistently picking a pool of $1000$ bit-strings out of the $10000$ samples obtained from the hardware, which get slightly less effective as the noise increases with problem size. On the other hand, the frequency filter leaves fewer and fewer samples to be used as warm starts as the noise increases, but as these are high quality samples, the classical algorithm benefits from them more.

We see a similar but more pronounced effect in the case of classical emulation, at least for up to an easily computationally tractable $N = 25$ nodes. In the cases of both the Hamming and frequency filters, only the solutions bit-strings remain after the filters are applied in many of the small-graph cases, leading to a maximum possible Q-factor across several graphs and even for varying number of QAOA layers $p$.

\subsubsection{SWAP-Enhanced heavy-hex graphs}\label{sec:enhanced_ibm_hw_results}

\begin{figure}
    \includegraphics[width=\textwidth]{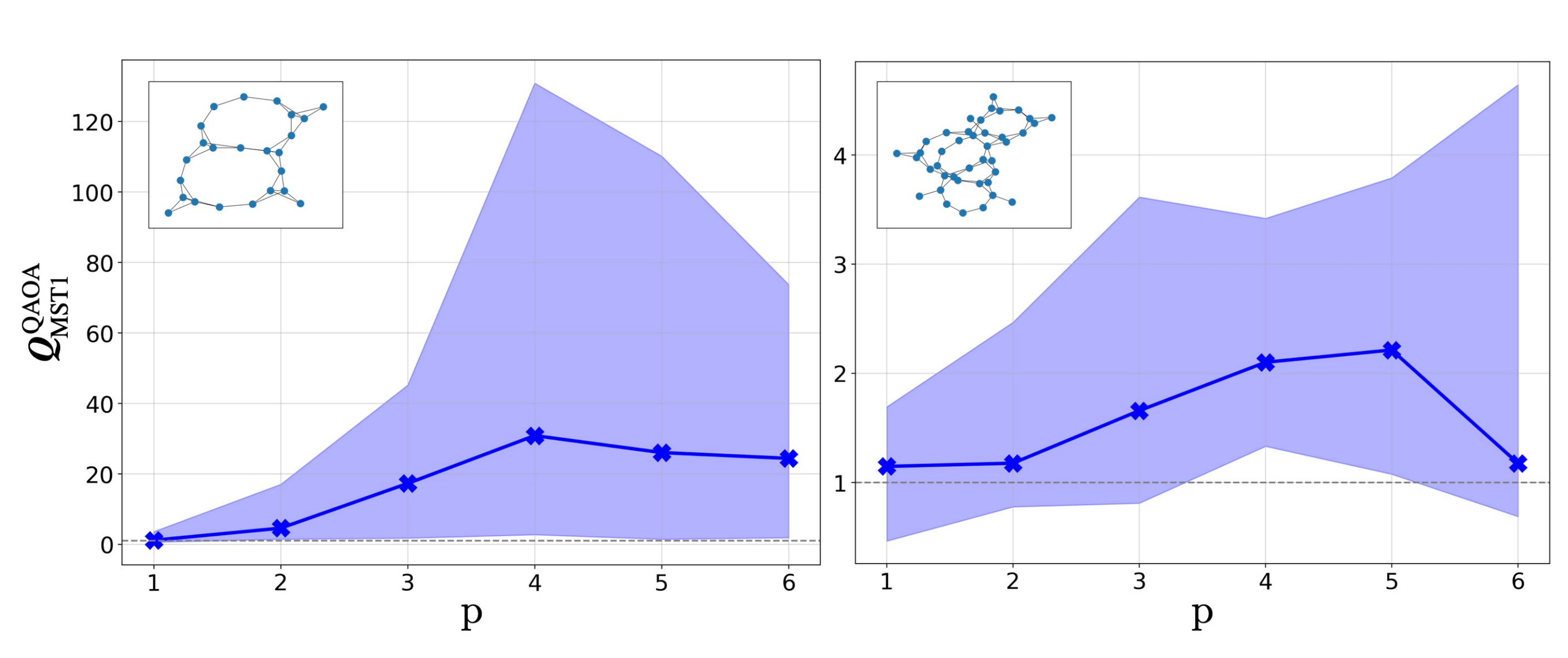}
    \caption{Plots of the Q-factor when using QAOA warm starts obtained from \texttt{ibmq\_torino} hardware with the Hamming filter and LNF readout error mitigation applied (see Sec.~\ref{sec:error_mitigation}) versus random bit-strings (PalubeckisMST1). In (a) we plot the median Q-factor across all $10$ of the 24 vertex graphs constructed as described in Appendix~\ref{sec:graphs_zz_swaps} while in (b) we plot the median result across all $10$ of the similarly enhanced 41-vertex graphs (see insets for the graph topologies). The Q-factor is obtained from 1000 optimizations each initialised either with a QAOA warm start or a random bit-string. The shaded regions indicate the largest and smallest speed-up obtained across all graphs for each number of QAOA layers $p$.}
    \label{fig:ibm_maxcut_swap_results_hamming}
\end{figure}

In addition to the line graphs, with unit weights on the edges, we also investigate Max-Cut instances of SWAP-enhanced graphs with the edge weights randomly sampled as follows: $w_{ij} \in [-1, 1]$. The SWAP-enhanced graphs are constructed as described in Appendix~\ref{sec:graphs_zz_swaps}, with edges added in an efficient way that reduces the total two-qubit gate count. We investigate $20$ graph instances: a set of $10$ graphs with $24$ vertices and $40$ edges with the same topology and different randomly generated sets of weights $W_i$ on the edges $\{G_{24}(V, E, W_{i})\}_{i = 1,...10}$ and a set of $10$ graphs with $41$ vertices and $68$ edges constructed in the same fashion: $\{G_{41}(V, E, W_{j})\}_{j = 1,...10}$. The insets in Fig.~\ref{fig:ibm_maxcut_swap_results_hamming} show the graph topologies of the two sets of graphs. 

In order to obtain the QAOA parameters for each of the graph instances, we use the angle-prediction methods from Sec.~\ref{sec:maxcut_angle_prediction}. We then implement the QAOA algorithm with predicted parameters on IBM's \texttt{ibmq\_torino} 133 qubit platform and post-process the bit-strings using the LNF readout error mitigation filter as well as the various classical filtering techniques described in Sec.~\ref{sec:em_filters}. We run 1000 optimizations with tabu search using either PalubeckisMST1 initialised with random bit-strings or warm-started with the outputs of the QAOA algorithm. 

We find that the Q-factor is largest when using the Hamming filter along with LNF readout error mitigation and we plot the results for this case in Fig.~\ref{fig:ibm_maxcut_swap_results_hamming} for both graphs. Unsurprisingly, we find that the the largest speedup of over $100\times$ is found for one of the smaller, $24$ vertex graphs, as the effects of decoherence are much larger in the $41$ vertex case. However, we still observe an almost $5$-fold speedup for some of the larger graphs, even in the case where $p = 6$ QAOA layers are used, requiring over $1000$ native two-qubit gates.

\begin{figure}
    \includegraphics[width=\textwidth]{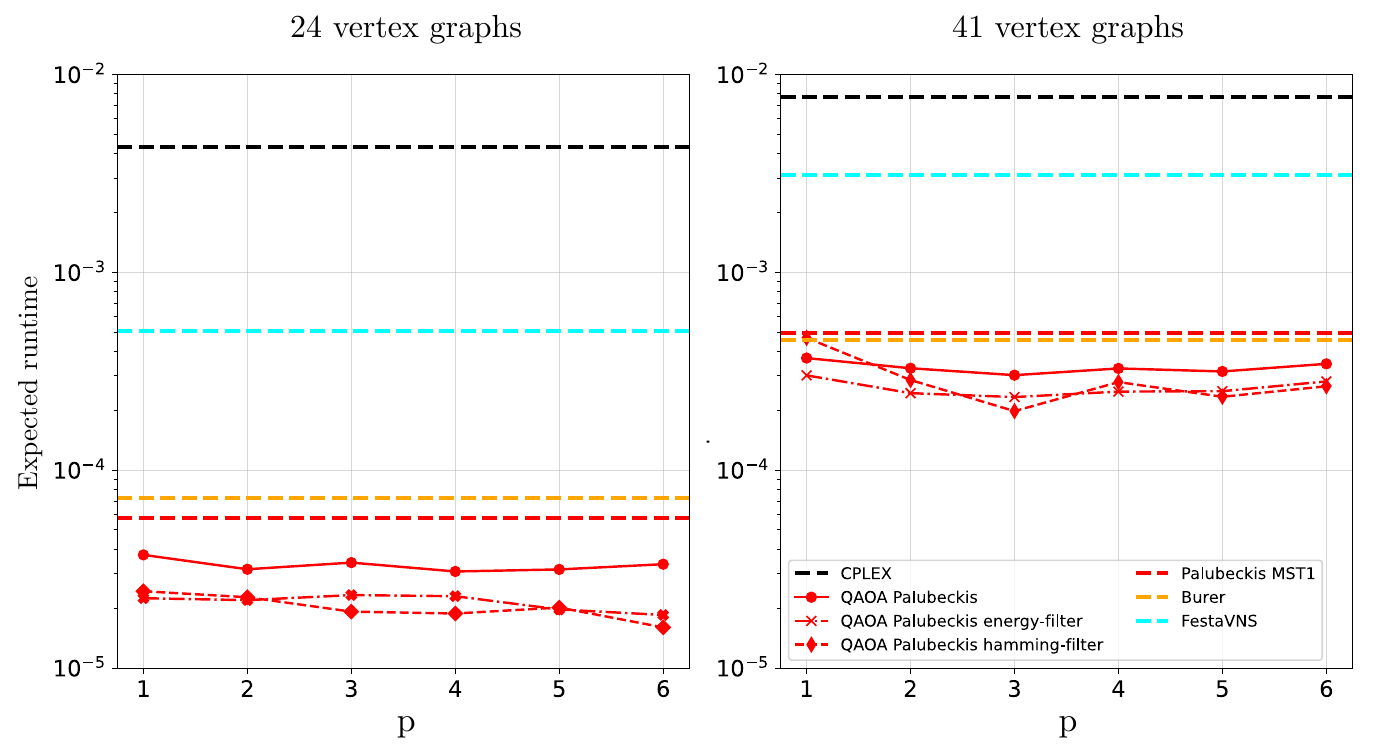}
    \caption{Comparing the expected runtimes of several of the best heuristics described in~\cite{dunning_what_2018} as well as CPLEX~\cite{cplex2009v12} to the QAOA warm-started version of PalubeckisMST1. We plot the median expected runtime (Eq.~\eqref{eq:expected_runtime}) across the 10 graph instanced in the 24 vertex case (left) and the 41 vertex case (right), as described in the main text.
    The horizontal dashed lines correspond to the expected runtimes of each of the classical heuristic algorithms, while the red plots correspond to expected runtimes when initializing PalubeckisMST1 with QAOA samples obtained from a run with $p$ QAOA layers, with their respective post-processing techniques. Note that these plots are comparing expected runtimes without the time required to generate/obtain the resource bit-strings for each of the heuristics. In the case of Burer, the timing is for the full runtime, as it does not get initialized with a random bit-string.}
    \label{fig:ibm_runtimes_combined}
\end{figure}

In order to compare our method to other heuristic algorithms as well as state-of-the-art solvers like CPLEX~\cite{cplex2009v12}, we run a \emph{modified} version of the Q-factor experiments where instead of running a single local search with a given warm start for a single optimization instance, we use the warm starts as a resource pool and run the local search algorithms several times during one optimization, re-starting from a different warm start every time. This means that the algorithm terminates not after a set number of iterations, but after some time limit, which in our case we set to $0.1$s. We begin timing the algorithm after generating the set of resource warm start and random bit-strings and then compute the expected runtime of Eq.~\eqref{eq:expected_runtime} using time in seconds instead of iterations $T$ as the variable. This captures a more direct notion of runtime, as not each iteration may take the same amount of time. 

In Fig.~\ref{fig:ibm_runtimes_combined}, we plot the median expected runtime across all $10$ graphs with both $24$ and $41$ vertices. We compare the expected runtime across $1000$ optimizations of the warm-started PalubeckisMST1 algorithm with energy-filtered and hamming-filtered QAOA samples with some of the best-performing classical heuristics from~\cite{dunning_what_2018} like Burer~\cite{burer_rank-two_2002}, FestaVNS~\cite{festa_randomized_2002} and PalubeckisMST1 as well as with the runtime of the state-of-the-art branch-and-bound based solver CPLEX~\cite{cplex2009v12}. We find that the QAOA warm-started tabu search converges faster across the graph instances than any of the classical methods, although the caveat is that this runtime comparison does not include the time taken to generate the warm starts or random samples in either case, meaning it is not a fair final comparison. However, this existing speed-up lends credence to the idea that for larger and harder problem instances, there may be a quantum advantage using our method, as the QAOA runtime for only a few layers is expected to grow far more slowly than the runtime of any classical method for increasing problem size. 

\begin{figure}[t]
\centering
    \includegraphics[width=\linewidth]{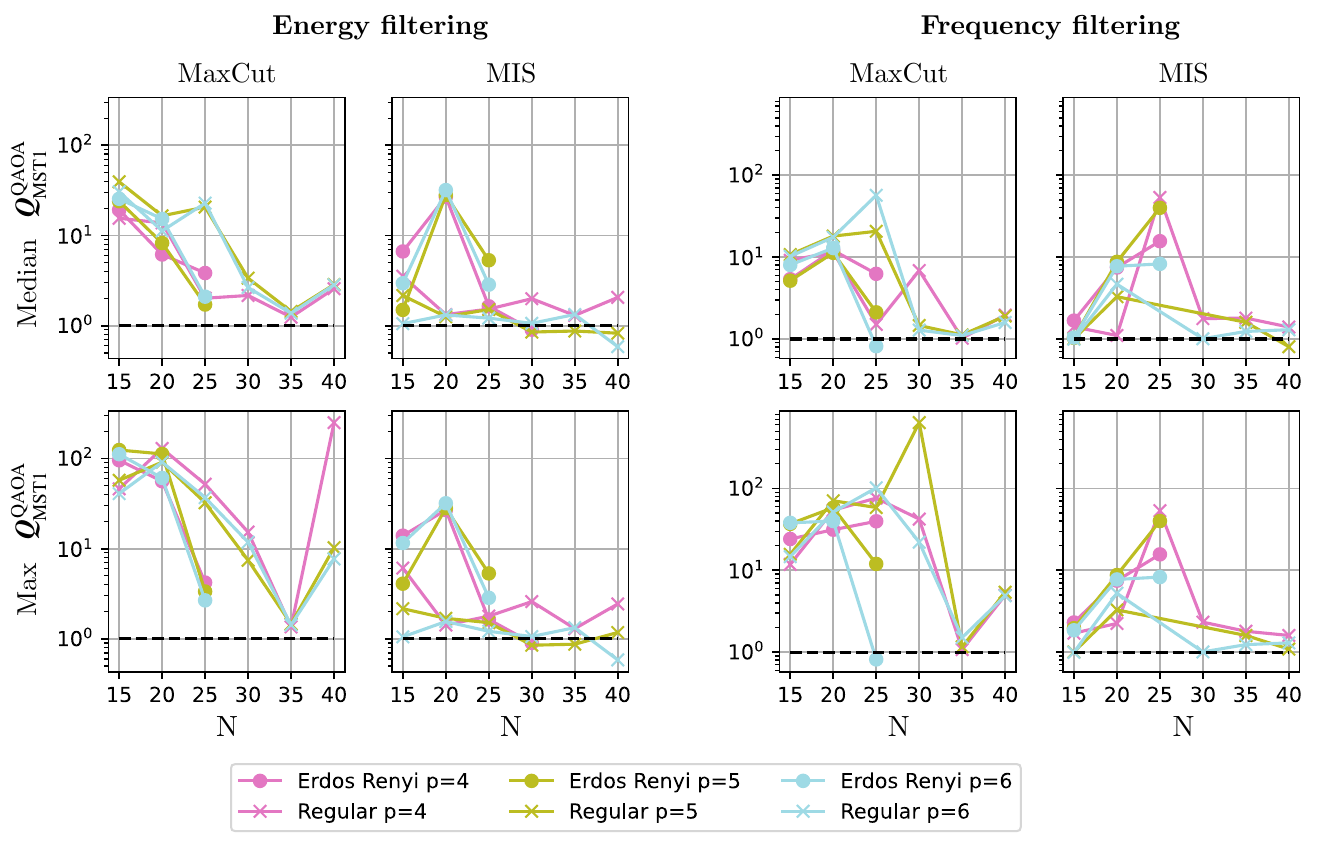}
\caption{The Q-factor from warm-starting the the PalubeckisMST1 algorithm implemented in MQLib~\cite{dunning_what_2018} with samples obtained from running the QAOA algorithm with predicted parameters using the `balanced' approach on the IBM \texttt{ibmq\_torino} chip, for Max-Cut and MIS on randomly generated Erd\H{o}s-R\'enyi graphs and regular graphs. Each point averages over graphs with edge probabilities between $0.1$ and $0.4$ for Erd\H{o}s-R\'enyi graphs and average degrees between $2$ and $5$ for random regular graphs. We plot the median Q-factor (top) and the maximum Q-factor (bottom) for the graphs with a given number of vertices. We see a speedup of up to a factor of $600$ using frequency filtering on a problem on $N=30$ nodes, and can still achieve a Q-factor of over $200$ for a problem on $N=40$ nodes. 
}
\label{fig:ibm_general_graphs}
\end{figure}

In Appendix~\ref{sec:appendix_further_hardware} we present further results with regards to the $41$ vertex graph instances.

\subsubsection{Erd\H{o}s-R\'enyi and random regular graphs}

We also test our algorithm on random instances of Max-Cut and MIS on Erd\H{o}s-R\'enyi graphs and regular graphs on between 15 and 40 nodes, with edge weights uniformly picked in $[1,-1]$ on IBM hardware using the \texttt{ibmq\_torino} chip. The edge probabilities for Erd\H{o}s-R\'enyi graphs range from $0.1$ to $0.4$, and the degrees for regular graphs range from $2$ to $5$. We use between 1 and 6 QAOA layers, not including any transpiled circuits that would have a single and 2-qubit circuit depth above 1000, and the QAOA angles were predicted using the `balanced' method from Section \ref{sec:maxcut_angle_prediction}. The problem graphs were mapped onto the hardware topology by using the quadratic assignment reduction for the initial qubit mapping and the A* routing method for the swap network construction described in Section \ref{sec:hardware}. For each problem instance, we took 10,000 shots on the hardware and used the energy filtering and frequency filtering methods to extract the best bit-strings to use as a warm start to the PalubeckisMST1 heuristic. 

In Figure \ref{fig:ibm_general_graphs}, we see the Q-factor obtained for these problem instances for $p=4,5,6$; the speedup for $p<4$ is much smaller as the ansatz is not expressive enough for these larger graphs. There is no data for the Erd\H{o}s-R\'enyi graphs on problems with more than $30$ nodes as the circuits were deeper than 1000 gates. In general, the Max-Cut problems perform better than the MIS problems, achieving a Q-factor of over $100$ in some cases. As the problem sizes get larger, we see noise from the device begin to take over as the circuit depth increases and we see $p=4$ outperforming $p=5$ and $p=6$ in the average case for $N=40$ instances. The frequency filter appears to outperform the energy filter above $N=20$ nodes in the problem, and in the best instances can achieve a Q-factor of nearly 600 on a $30$ vertex graph. We see that the performance of the quantum enhanced optimization algorithm highly depends on the problem instance and also the samples we get from the device, where the best instances can have a Q-factor 1 or even 2 orders of magnitude higher than the median case. Compared to Figure \ref{fig:palubeckis_er_0.3_reg_4} which shows the results for unweighted random regular and Erd\H{o}s-R\'enyi graphs with similar densities, for the graphs with under 20 vertices we see a slightly worse performance relative to the unfiltered noiseless emulation results.  


\section{Application to Arbitrarily Large Problems}\label{sec:division}

While in this paper we focus on demonstrating quantum speedups for problem instances that are small enough to be solved directly on real or emulated quantum hardware, larger instances can also be addressed using our approach. There are multiple ways of dividing a large optimization problem into subproblems: for example, recursive/hierarchical decomposition, partial expansion of a backtracking tree, and simply dividing the input graph into subgraphs. The last of these is a natural fit to our quantum-enhanced approach.
A classical clustering algorithm can be used to divide up the graph to minimize connections between subgraphs; in our \href{https://optimization.phasecraft.io}{software tool}, we used the METIS package~\cite{metis}. Then a solution to the overall problem can be found by simply concatenating the returned solutions to the subproblems. However, in the case of the Max-Cut problem, this may not lead to an high-quality answer immediately. This is because of the overall $\Z_2$ symmetry of Max-Cut: for each solution to a subproblem we can get an equally good solution to that subproblem by flipping all the bits (and QAOA will choose each of the two solutions with equal probability), yet the quality of an overall solution built from combining solutions to multiple subproblems depends on the choice of bit flips for each one.

To produce a low-cost solution to the overall problem, one can optimize over all flips to the bits of subproblems. This optimization problem turns out to correspond to a QUBO instance on an effective graph whose vertices correspond to subproblems, with weighted edges whose weights are determined by the edges between subproblems. It was proposed in~\cite{zhou23} that this overall QUBO problem could be solved itself by using QAOA. Here, we instead observe that in general this problem will be small enough that it can be solved exactly and efficiently using a classical heuristic for Max-Cut or QUBO problems. Combining these ingredients enables instances on hundreds or thousands of vertices to be solved using today's quantum hardware.


\section{Discussion and Conclusions}
\label{sec:discussion}

This work establishes a framework for leveraging QAOA-generated samples to accelerate classical optimization. Our method demonstrates that even in the NISQ era, where quantum devices are constrained by noise and limited qubit connectivity, a quantum-assisted strategy can provide measurable performance benefits. Through a combination of novel parameter-setting techniques, hardware-aware circuit optimization and error-mitigation strategies, we have systematically improved the integration of QAOA within classical optimization pipelines. We demonstrated the effectiveness of our scheme both in emulation and by running on real noisy quantum hardware, showing that we can expect a speedup in the convergence time of classical heuristics even when using samples from deep circuits with over $1000$ two-qubit gates.

\pagebreak

Our experiments highlight several key findings. First: the introduction of quantum warm starts significantly improves the performance of classical heuristics like tabu search in instances where classical optimization algorithms rely on random initializations. The speedup, as measured by the Q-factor, reaches values exceeding $1000\times$ in certain cases, with performance gains persisting across different problem sizes. Second: despite the hardware constraints, we demonstrated that customized qubit mapping and routing strategies (e.g., A* heuristic and greedy approaches) reduce circuit depth, making QAOA implementations more feasible. Our SWAP-optimized graph encoding allowed us to extend the reach of QAOA on real quantum processors while maintaining low error rates. Third: the proposed parameter heuristics from Sec.~\ref{sec:parameters} offer a scalable alternative to classical variational optimization, providing near-optimal QAOA parameters without computationally expensive training. The validity of these parameter-setting techniques was supported through their consistent performance across both Max-Cut and MIS instances. And finally: by integrating readout-error mitigation (LNF filter) and classical bit-string filtering techniques (energy-filtering, frequency-filtering, and Hamming filtering), we substantially improved the quality of QAOA samples with simple and very efficient classical post-processing techniques. 

Despite this, several questions remain to be answered before this approach can be said to have delivered a genuine quantum advantage. The first concerns itself with problem size: our study focused on problem sizes up to $55$ qubits on IBM hardware. Future research should explore whether the observed speedups persist or improve when partitioning and recombining larger-scale instances. Additionally, investigating QAOA’s behaviour on denser and more complex combinatorial problems (e.g.\@, industrial scheduling problems, protein folding, or financial portfolio optimization) would provide valuable insights. Are there problem instances that QAOA is particularly suited to speed-up due to limits of classical local search or greedy initialisations?

Secondly, although our results indicate a substantial improvement over randomly initialized heuristics, further work is needed to benchmark our method against the best-in-class classical solvers on larger-scale problem instances. The question remains: can this approach surpass state-of-the-art classical heuristics like Burer’s method~\cite{burer_rank-two_2002} or outperform advanced mixed-integer programming solvers like Gurobi~\cite{gurobi} or CPLEX~\cite{cplex2009v12} for harder problem instances? Our runtime metrics did not take into account the runtime required to obtain the QAOA samples and it is likely that we will require testing on far larger problem instances to see a runtime improvement where the full pipeline is taken into account. While in Appendix~\ref{sec:classical_benchmarks} we explore the performance of such classical methods on a few structured families of problems, it is hard to evaluate what problem instances are truly difficult across the board for any single algorithm, much less for all of them. 

Finally, our results suggest that as hardware fidelity and qubit count increase, quantum-enhanced optimization could become increasingly viable. As gate errors decrease and coherence times improve, higher-depth QAOA circuits ($p > 6$) may become practical, potentially leading to an even more pronounced quantum speedup. Furthermore, larger problem instances will become accessible as higher-depth circuits will not suffer from as much decoherence. As hardware improves, the viability of quantum advantage using our scheme will improve with it.

To conclude, our results provide compelling evidence that even with today’s noisy quantum processors, QAOA-generated samples can accelerate classical optimization. By leveraging structured quantum sampling, problem-aware circuit optimizations, and efficient parameter-setting techniques, we achieve significant speedups over traditional heuristics. This suggests that hybrid quantum-classical approaches can provide practical benefits long before fully fault-tolerant quantum computers become available. We remark that it would also be possible to experiment with using advanced methods for simulating QAOA, such as the recently proposed noisy sampling method based on Pauli Propagation of Ref.~\cite{martinez2025sampling} to replace the quantum part of the algorithm, and produce a quantum-inspired warm start method. While the approach only approximates the output of QAOA circuits and is likely not to capture the correlations that relatively low-noise hardware could, it may offer an insight into the structure of the distributions that noisy near-term QAOA can produce.

With further improvements in quantum hardware and more sophisticated hybrid strategies, we anticipate that quantum-enhanced optimization will become an essential tool for tackling large-scale combinatorial problems in various industries, from logistics and finance to power grid optimization.

\section*{Acknowledgments}
We would like to thank Sami Boulebnane and Pete Rolph for helpful discussions on the topic of this paper, and Ra\'ul Garc\'ia-Patr\'on for comments on an earlier version. This report is work commissioned by Innovate UK. We also acknowledge the use of IBM Quantum services for this work. The views expressed in this publication are those of the authors and not necessarily those of Innovate UK, IBM or the IBM Quantum team. AM and LZ acknowledge funding from the European Research
Council (ERC) under the European Union’s Horizon 2020 research and innovation programme (grant agreement No.\ 817581).

\bibliographystyle{mybibstyle}
\bibliography{qeopt}

\appendix

\section{Predicting near-optimal QAOA parameters}\label{sec:appendix_predicting}

\begin{figure}[t]
    \centering
    \includegraphics[width=\textwidth]{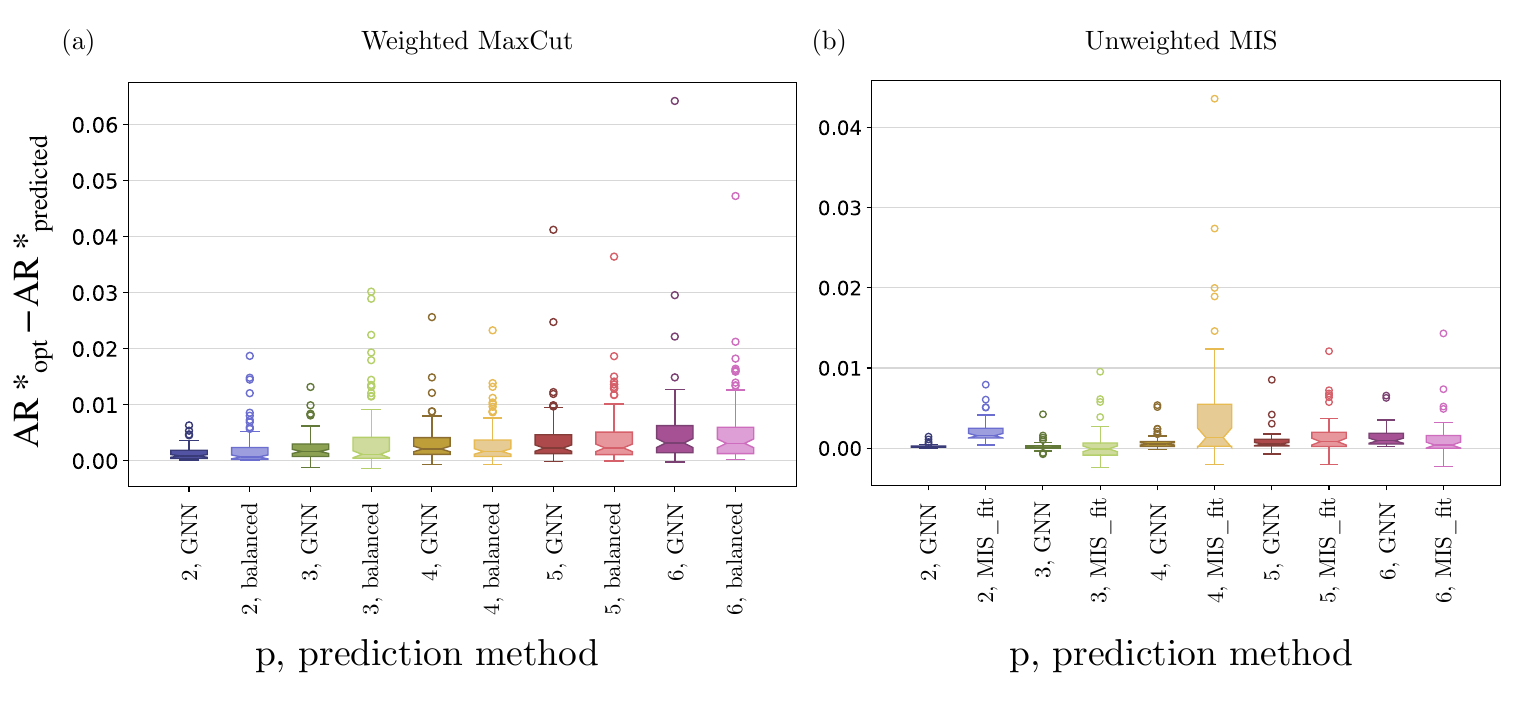}
    \caption{Performance of GNN model as compared to previous methods in predicting optimal parameters for (a) Max-Cut instances with weights sampled randomly $\in [-1,1]$ as well as (b) the unweighted MIS case with penalty $\lambda = 1$. The plots show the difference in modified approximation ratio AR* for states that use optimized QAOA angles ($\rm{AR*}_{\rm opt}$) versus those that have been predicted using either GNNs or the ``balanced" method in the case of Max-Cut (Sec.~\ref{sec:maxcut_angle_prediction}), as well as the unweighted MIS method (MIS\_fit) presented in Sec.~\ref{sec:angle_prediction_MIS}. }
    \label{fig:gnn_results_boxplots} 
\end{figure}

This section expands on the details and performance of our parameter prediction methods for QAOA. We are primarily interested in examining the performance of prediction methods on problem instances which are larger than those used to formulate the method or train the predictions, but still small enough to allow for classical optimization in order to test their performance. In Fig.~\ref{fig:gnn_results_boxplots} (a) we plot a comparison of the ``balanced'' approach (Eq.~\eqref{eq:balanced_method}) for predicting weighted Max-Cut parameters to GNN model predictions, for 90 Erd\H{o}s-R\'enyi graphs with $N \in [21, 25]$ nodes. The edge weights are generated randomly from a uniform distribution in the range $[-1,1]$ and the edge probabilities range from $0.1$ up to $0.9$. The plot shows the difference in AR* metric from Eq.~\eqref{eq:approx_ratio_rescaled}, between the state obtained using predicted parameters and those obtained using classical optimization. The results show that both prediction methods lead to a very small error in the final energy of the state as compared to optimization, even for $p = 6$.

In Fig.~\ref{fig:gnn_results_boxplots} (b), we plot a similar comparison of the MIS fitting method (``MIS\_fit'') and GNN model predictions, for unweighted MIS graphs. The graph sizes and structures are generated in the same manner as in the Max-Cut case, but the QUBO weights are determined by the fact that we are investigating unweighted MIS instances. We find that both the 

\subsection{Graph Neural Networks for parameter prediction}\label{sec:appendix_predicting_gnns}

Graph neural networks (GNNs) are a family of neural networks that can operate naturally on graph-structured data. This is a consequence of their ability to learn functions on graphs that are invariant under graph isomorphisms and to preserve node-order equivariance throughout training. In the case of QAOA parameters, node relabeling should not lead to changes in the optimal parameters, since they act globally on the entire graph, thus we require that any function representing a relationship between a QUBO graph and its QAOA parameters be invariant to graph isomorphisms. We further observe in both the Max-Cut and MIS prediction methods from Secs.~\ref{sec:maxcut_angle_prediction} and \ref{sec:angle_prediction_MIS}, that the core patterns in optimal QAOA parameters can be captured by global properties of the problem graph, like its average degree, indicating that machine learning methods like GNNs should be suitable for capturing such patterns. This is a consequence of ML methods dealing better with high-dimensional data which can be well-approximated by a low-dimensional projection (e.g.~statistical properties like average degree or average edge weight)~\cite{bronstein2021geometric}.

In our approach, we train a separate GNN model for each number of QAOA layers, matching the number of convolutional layers to the number of QAOA layers in each case. We find that the best-performing architecture for our purpose is the edge-aware Graph Isomorphism Network (GINEConv \cite{hu2020strategies}), as implemented in \texttt{Pytorch-Geometric} \cite{fey2019fast}, where at the $k^{\rm th}$ step of training, each layer updates node representation $h_v$ of the $v^{\rm th}$ node as follows:
\[
h_v^{(k)} = \text{MLP}\left((1 + \epsilon^{(k)})h_v^{(k -1)} + \sum_{u \in \mathcal{N}(v)} \text{ReLU}(h_u^{(k-1)} + e_{uv})\right).
\]
where $e_{uv}$ are the edge weights of the edges that lead to the neighboring nodes of $v$ and \text{MLP} is a multilayer perceptron - a stack of linear and non-linear neural network layers.

Together with a global pooling layer and an output prediction layer with dimension $2p$ for all of the required QAOA parameters, each model takes as input a representation of a QUBO graph, with node and edge weights corresponding to $c$ and $Q$ from Eq.~\eqref{eq:qubo} respectively, and outputs its predicted near-optimal QAOA parameters. Our training data consists of Erd\H{o}s-R\'enyi QUBO graphs with $N \in [10, 20]$ nodes and edge probabilities $p_E$ sampled to produce both sparse and dense examples $p_E \in [0.1, 0.9]$. The node and edge weights are all scaled to be in the range $[-1,1]$, with various graph problem structures included in the training set: cases where all node and edge weights are uniformly randomly sampled, cases with fixed node/edge weights and uniformly sampled edge/node weights respectively, as well as instances of Max-Cut and MIS. Each training data point is a graph coupled with its optimized QAOA parameters, where the optimization is performed using exact-diagonalisation emulations of the circuits.

We find that for each GNN model, the amount of training data for good performance is relatively low -- in the range of several thousand graphs of various sizes and problem types -- and there is evidence of such an approach scaling well with increasing problem size. For example, in Fig.~\ref{fig:gnn_results_boxplots}, the results show that for both Max-Cut and MIS QUBO instances, the error in energy compared to optimized parameters is competitive with our other prediction methods, despite the test instances being larger than those found in training. 

\subsubsection{Previous work}

This is not the first attempt to implement GNNs for learning the structure of QAOA parameters: in \cite{deshpande_capturing_2022}, the authors implement effectively the same approach, although it is studied solely for the Max-Cut case and for very small graph instances. Similarly, a more recent work \cite{liang_invited_2024} focuses on using GNNs to predict near-optimal QAOA parameters for Max-Cut in order to be used as warm starts for online optimization. Finally, a similar approach is used for MIS problem instances in \cite{xu2025qaoa}. These works all follow from the intuition that GNNs are a machine learning model which should be well-suited to capturing the structure of QAOA parameters based solely on the graph structure of each QUBO, similar to how our methods for Max-Cut and MIS above use global properties of each QUBO graph in their prediction strategies. We follow this intuition, extending it to general QUBO problems beyond Max-Cut and MIS.

\section{SWAP-enhanced heavy-hex graphs}\label{sec:graphs_zz_swaps}

In order to reduce quantum resource requirements when implementing more interesting circuits on the IBM superconducting hardware, the graph problem instances can be tailored to match the connectivity of the devices they are implemented on. Given the heavy-hex connectivity graphs of the IBM devices $G_{\rm IBM}(V,E)$, we generate problem instances of the form $G'(V', E', W)$ where $G'$ is an induced subset of $G_{\rm IBM}$ with weighted edges according to the set of weights $W$ and $V' \subseteq V$.

This method of generating graphs removes the requirement for qubit swapping, as each edge in the problem graph corresponds to a physical coupling on the device. However, this also leads to relatively sparse graphs with simple structure which does not allow us to test our scheme on more interesting problem instances. 

In order to enhance the problem instances without increasing the two-qubit gate counts too much, we combine the $ZZ$ interactions in the cost Hamiltonians of Eq.~\eqref{eq:maxcut_hamiltonian} and Eq.~\eqref{eq:mis_hamiltonian} with the necessary SWAP interactions which allow for additional edges to be added to the graph instances while requiring less total two-qubit gates. If one wishes to perform a $ZZ$ interaction between qubits $i$ and $j$ corresponding to an edge $(i,j) \in E$ of the problem graph, followed by a swap of qubits $i$ and $j$, one can simplify the entire operation as follows:
\begin{center}
\begin{tikzpicture}
   \begin{yquantgroup}
      \registers{
         qubit {} q[3];
      }
      \circuit{
         cnot q[1] | q[0];
         box {$RZ(w_{ij}\gamma)$} q[1];
         cnot q[1] | q[0];
         cnot q[1] | q[0];
         cnot q[0] | q[1];
         cnot q[1] | q[0];
      }
      \equals
      \circuit{
         cnot q[1] | q[0];
         box {$RZ(w_{ij}\gamma)$} q[1];
         cnot q[0] | q[1];
         cnot q[1] | q[0];
      }
   \end{yquantgroup}
\end{tikzpicture}
\end{center}
where the LHS circuit removes 2 unnecessary CNOT operations, which on the IBM hardware is equivalent to 4 native $CZ$ two-qubit operations. This is a simple reduction, but an additional trick allows for the efficient construction of problem instances with added edges between vertices that are a distance of two edges away from each other, leading to a far richer set of problem instances to test on. The method for doing this is as follows:
\begin{enumerate}
    \item Given a subgraph of the device graph $G'(V', E', W)$, select a set of edges $E_{\rm SWAP} \subset E'$ which will act as SWAP edges, allowing each vertex in the edge to access the neighbouring vertices of the other.
    \item Generate a set of new edges $E_{\rm new}$ as follows: for each $(i,j) \in E_{\rm SWAP}$, add $\{(i, j')\}_{j' \in \mathcal{N}(j)\setminus \{i\}}$ and $\{(i', j)\}_{i' \in \mathcal{N}(i)\setminus \{j\}}$ to $E_{\rm new}$, where $\mathcal{N}(i)$ is the set of all neighbouring vertices of $i$. We assume that for every vertex $i$ incident to an edge in $E_{\rm SWAP}$, all neighbours $\mathcal{N}(i) \setminus \{j\}$ are not incident to an edge in $E_{\rm SWAP}$.
    \item When implementing QAOA, for odd number of layers, first implement the $ZZ$ interactions corresponding to the edges $E'\setminus E_{\rm SWAP}$, followed by a set of combined $ZZ$ and SWAP interactions for each of the edges in $E_{\rm SWAP}$. Finally, perform a $ZZ$ interaction corresponding to each new edge in $E_{\rm new}$.
    \item In the case where the number of layers is even, perform the three steps in reverse, such that the qubit information ends up back in the original physical qubits that it began in at the start of the algorithm.
    \item If the QAOA algorithm is run for total $p$ which is odd, instead of swapping back the qubits at the end, keep track of the SWAPs and perform post-processing on the output bit-strings to recover the correct measurements corresponding to the un-SWAPped qubits.
\end{enumerate}

\section{Further Emulation Results}\label{sec:appendix_further_emulation}

\begin{figure}[t]
\centering
    \includegraphics[width=0.78\linewidth]{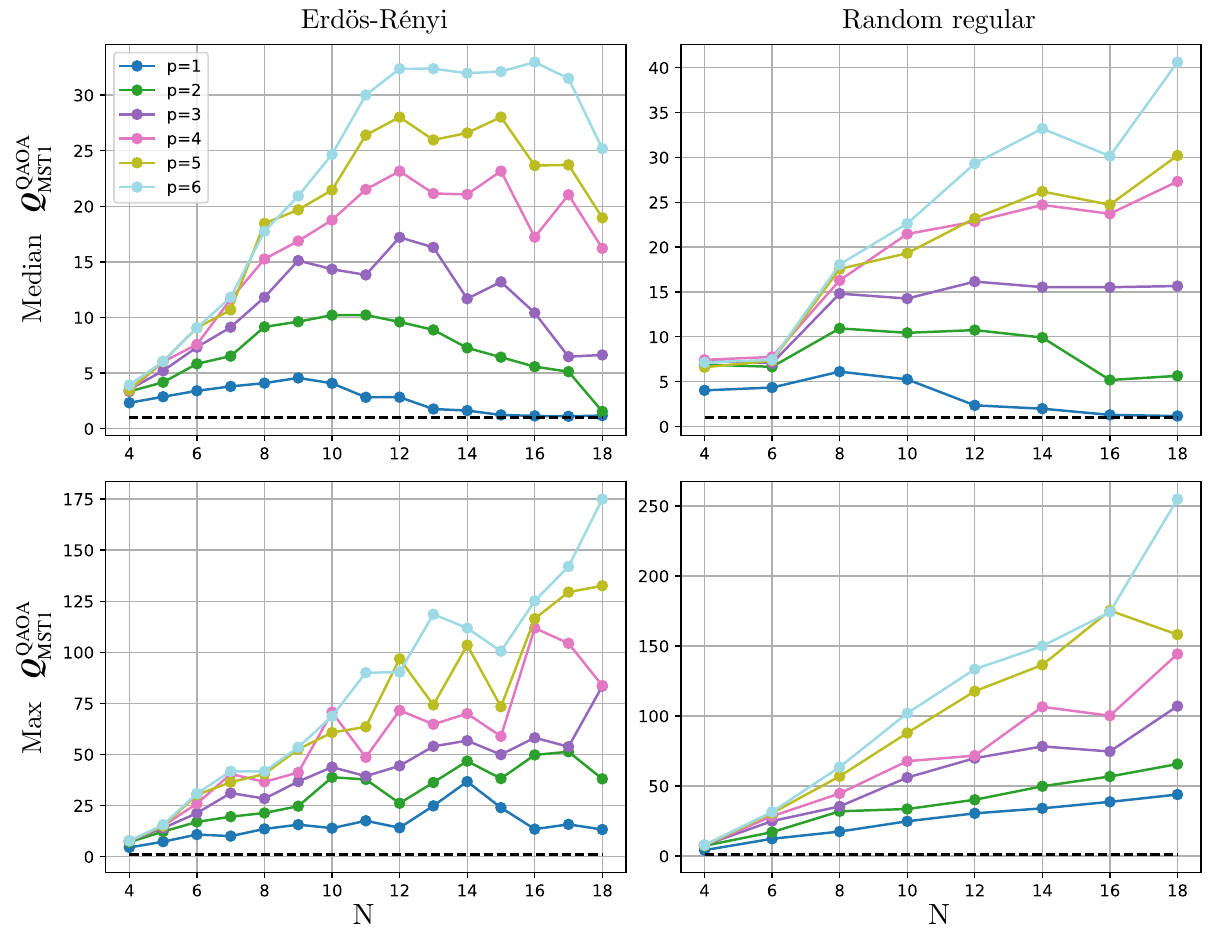}
\caption{The Q-factor from warm-starting the the PalubeckisMST1 algorithm implemented in MQLib~\cite{dunning_what_2018} for unweighted Max-Cut with samples obtained using classical emulation of the QAOA algorithm for randomly generated Erd\H{o}s-R\'enyi graphs (left) and regular graphs (right). Each point averages over 10 graphs for every edge probability between $0.1$ and $0.9$ for Erd\H{o}s-R\'enyi graphs and every average degree between $1$ and $N-2$ for random regular graphs. We plot the average Q-factor (top) and the maximum Q-factor over any graph of a given size (bottom). For $p=6$ we see a Q-factor of up to $175$ for Erd\H{o}s-R\'enyi graphs and $250$ for random regular graphs, and on average a Q-factor of $40$ and $60$ respectively.
}
\label{fig:emulated_random_graph_speedups_palubeckis}
\end{figure}

We evaluated the performance of the quantum enhanced optimization algorithm on a large set of randomly generated unweighted graphs. We explored both randomly generated Erd\H{o}s-R\'enyi graphs with edge probabilities between $0.1$ and $0.9$ and random regular graphs with average degrees between $1$ and $N-2$ where $4 \leq N \leq 19$ is the number of nodes in the problem graph. For each parameter and graph size $N$ we generate $10$ random graphs. We use the `balanced' approach from Sec.~ \ref{sec:maxcut_angle_prediction} to predict the QAOA parameters. We obtain $1000$ samples from the classical emulator, and use these as a warm start to either the PalubeckisMST1 or the FESTA2002VNSPR heuristics implemented in MQLib~\cite{dunning_what_2018}. We chose these as they were some of the highest performing heuristics in MQLib (Table 1 in~\cite{dunning_what_2018}) that use a random initial bit-string to start a local search, and so are amenable to a warm start.  Finally we estimate the speedup in Figure \ref{fig:emulated_random_graph_speedups_palubeckis} using the Q-factor from Eq.~\eqref{eq:qfactor}.

\subsection{PalubeckisMST1}

 Averaging over all graphs of a given size, for the PalubeckisMST1 algorithm~\cite{palubeckis_multistart_2004} we see a speedup of up to a factor of $35$ on Erd\H{o}s-R\'enyi graphs and up to $40$ on random regular graphs in Figure \ref{fig:emulated_random_graph_speedups_palubeckis}. When looking at the problem instance that provides the largest speedup, we can achieve a Q-factor of over $250$ for a random regular graph with $18$ vertices. We see a better performance with the random regular graphs as the balanced approach for predicting angles is based on graphs of constant degree. There is also a positive scaling as we increase both the number of vertices in the problem graph up to $N=18$ and increase the number of layers of QAOA up to $p=6$. For small number of layers $p$, we begin to see a decrease in performance as $N$ increases, especially for Erd\H{o}s-R\'enyi graphs, likely because the ansatz is not expressive enough and so is not returning useful bit-strings to use as a warm start. 

\begin{figure}[t]
\centering
    \includegraphics[width=0.9\linewidth]{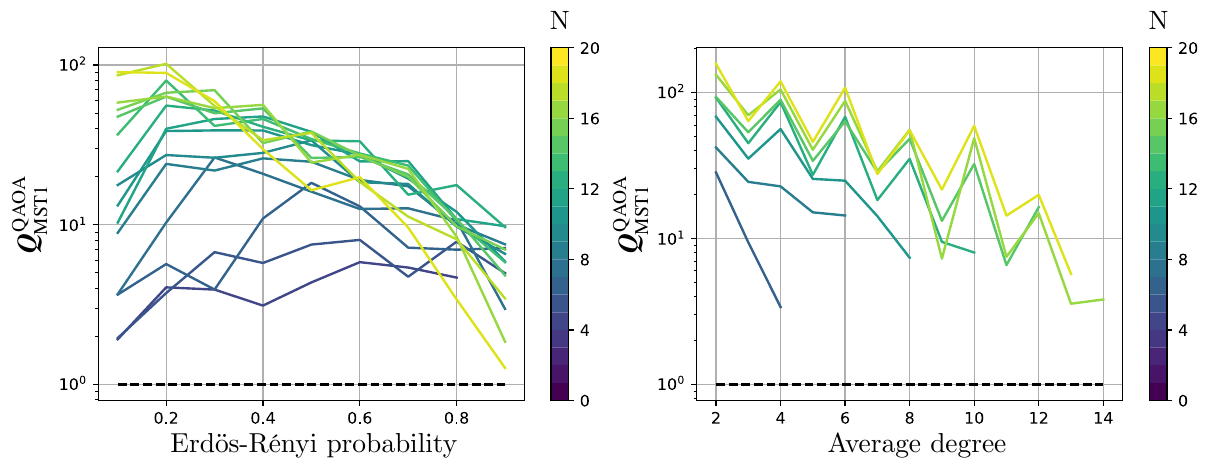}
\caption{The Q-factor from warm-starting the the PalubeckisMST1 algorithm using $p=6$ QAOA with the same set of Erd\H{o}s-R\'enyi graphs (left) and regular graphs (right) as Figure \ref{fig:emulated_random_graph_speedups_palubeckis}, as a function of the density of the graph. Each point averages over 10 graphs for unweighted Max-Cut. In general, the denser the graph is the smaller the speedup that is obtained.
}
\label{fig:palubeckis_density_speedup}
\end{figure}

\begin{figure}[t]
\centering
    \includegraphics[width=0.9\linewidth]{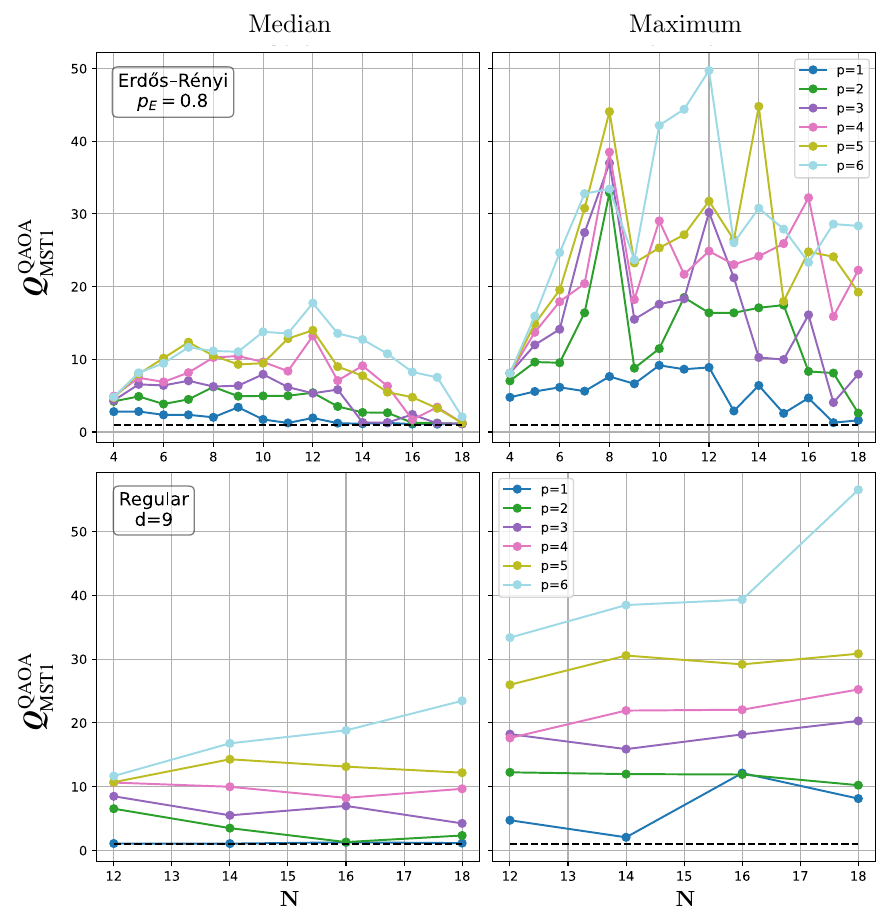}
\caption{The Q-factor from warm-starting the the PalubeckisMST1 algorithm implemented in MQLib~\cite{dunning_what_2018} for unweighted Max-Cut with samples obtained using classical emulation of the QAOA algorithm using the `balanced' approach for predicting angles for 10 randomly generated Erd\H{o}s-R\'enyi graphs for edge probability $p=0.8$ (top) and 10 regular graphs with degree $d=9$ (bottom). These are denser problem instances compared to Figure \ref{fig:palubeckis_er_0.3_reg_4}, and we see a much worse performance only achieving a median Q-factor of around 20. 
}
\label{fig:palubeckis_dense_graphs}
\end{figure}

The density of the graph in the problem instance is set by the edge probability for Erd\H{o}s-R\'enyi graphs and degrees for random regular graphs. In Figure \ref{fig:palubeckis_density_speedup} for $p=6$ QAOA layers we see that as the density increases, the Q-factor decreases on average for both sets of problems. We also see this in Figure \ref{fig:palubeckis_dense_graphs}, where compared to Figure \ref{fig:palubeckis_er_0.3_reg_4} the performance is much worse and no longer significantly increases as the problem size increases. This could be due to a variety of reasons: the angle prediction methods may not perform as well for denser graphs, the quantum algorithm may require more layers to provide similar quality samples, or the classical heuristic may not benefit as much from a warm start. For small Erd\H{o}s-R\'enyi graphs, there is a slight increase in performance as the density increases, which may be because small sparse problems are too easy for the classical heuristic to solve, and so warm-starting does not provide much benefit. Even degree random regular graphs also appear to perform better than odd degree graphs across all instance sizes. The Q-factor varies much more for denser graphs, shown by the larger difference between the median and maximum Q-factors in Figure \ref{fig:palubeckis_dense_graphs} and chaotic nature of the maximum Q-factor, indicating the performance is highly dependent on the structure of the problem graph. The scaling with $N$ is less evident in this case, where as the problem size grows we no longer see a consistent increase the Q-factor.

\subsection{FESTA2002VNSPR}\label{sec:festa_results}

\begin{figure}[t]
\centering
    \includegraphics[width=0.82\linewidth]{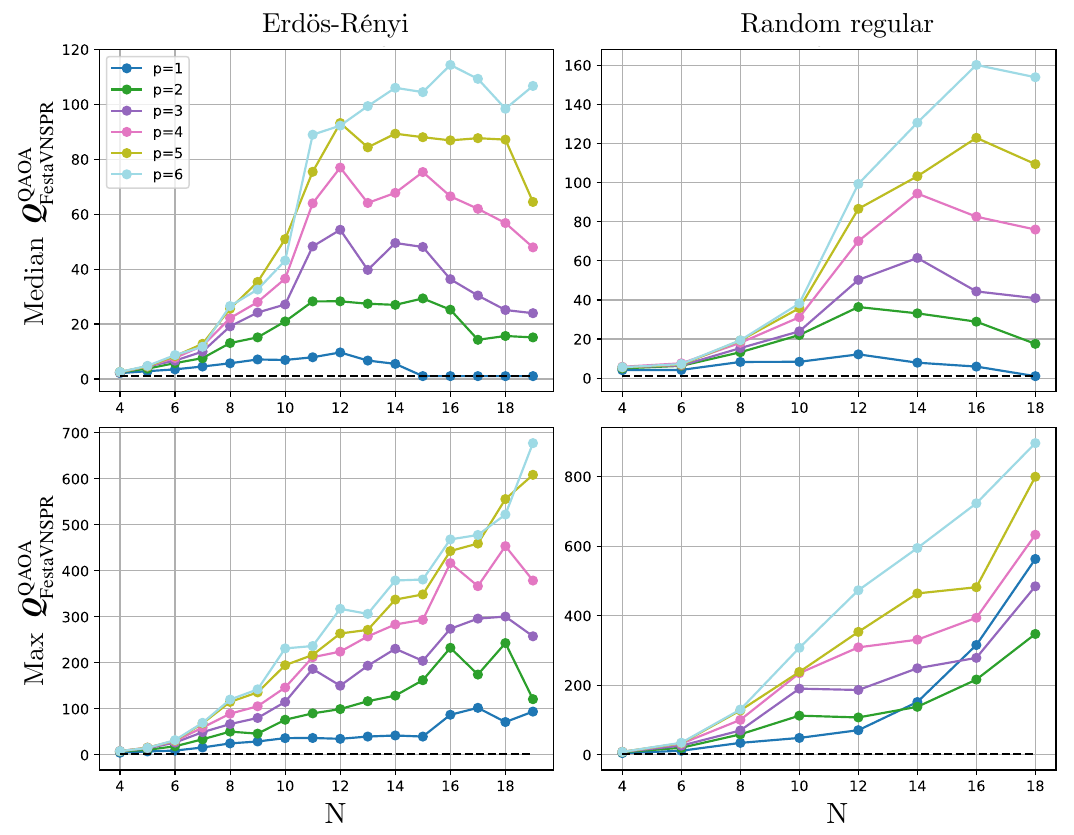}
\caption{The Q-factor from warm-starting the the FESTA2002VNSPR algorithm implemented in MQLib~\cite{dunning_what_2018} for unweighted Max-Cut with samples obtained using classical emulation of the QAOA algorithm using the `balanced' approach for predicting angles for randomly generated Erd\H{o}s-R\'enyi graphs (left) and regular graphs (right). Each point averages over 10 graphs for every edge probability between $0.1$ and $0.9$ for Erd\H{o}s-R\'enyi graphs and every average degree between $1$ and $N-2$ for random regular graphs. We plot the average Q-factor (top) and the maximum Q-factor over any graph of a given size (bottom). We see a similar qualitative behaviour to Figure \ref{fig:emulated_random_graph_speedups_palubeckis}, with increasing Q-factor as $N$ increases and the number of layers $p$ increases. The regular graphs also perform better than the Erd\H{o}s-R\'enyi graphs, where we see a median Q-factor of up to $160$ and maximum Q-factor of up to $900$.}
\label{fig:emulated_random_graph_speedups_festa}
\end{figure}

\begin{figure}[t]
\centering
    \includegraphics[width=\linewidth]{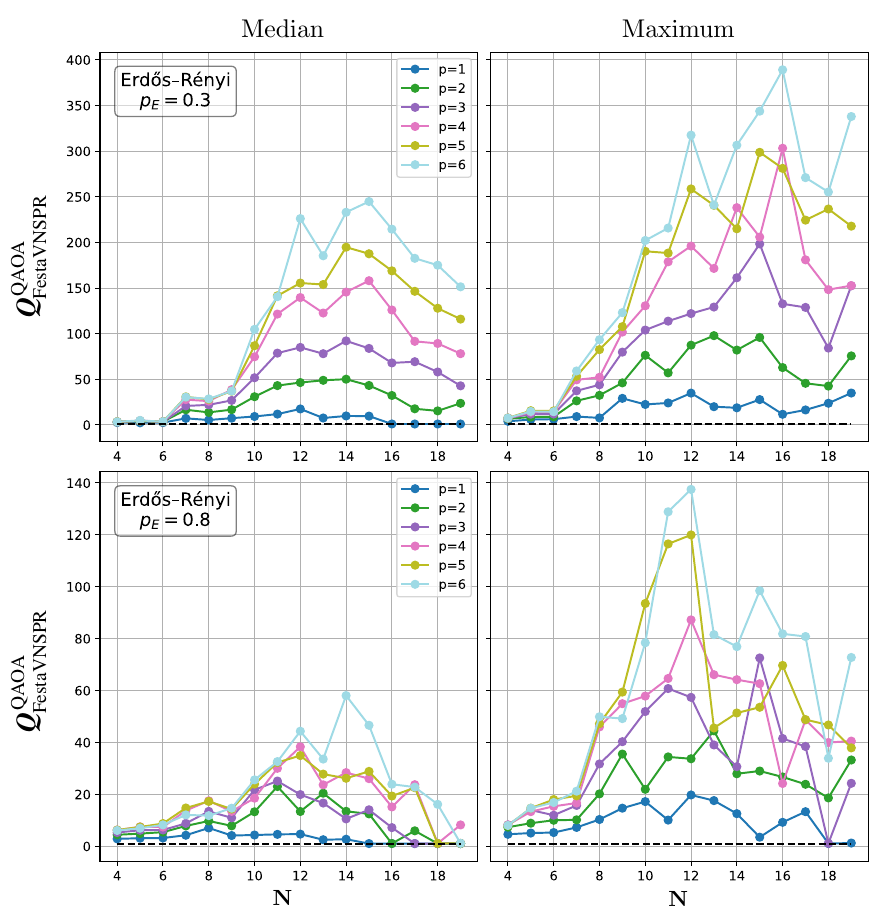}
\caption{The Q-factor from warm-starting the the FESTA2002VNSPR algorithm implemented in MQLib~\cite{dunning_what_2018} for unweighted Max-Cut with samples obtained using classical emulation of the QAOA algorithm using the `balanced' approach for predicting angles for 10 randomly generated Erd\H{o}s-R\'enyi graphs with edge probability $p=0.3$ (top) and 10 graphs with $p=0.8$ (bottom). We see a much better performance in both the average and maximum cases for less dense graphs, where for $p=6$ layers we can achieve median Q-factors of up to 250 for graphs with $p=3$, whereas only up to 60 for graphs with $p=0.8$.}
\label{fig:festa_er_different_p}
\end{figure}

\begin{figure}[t]
\centering
    \includegraphics[width=\linewidth]{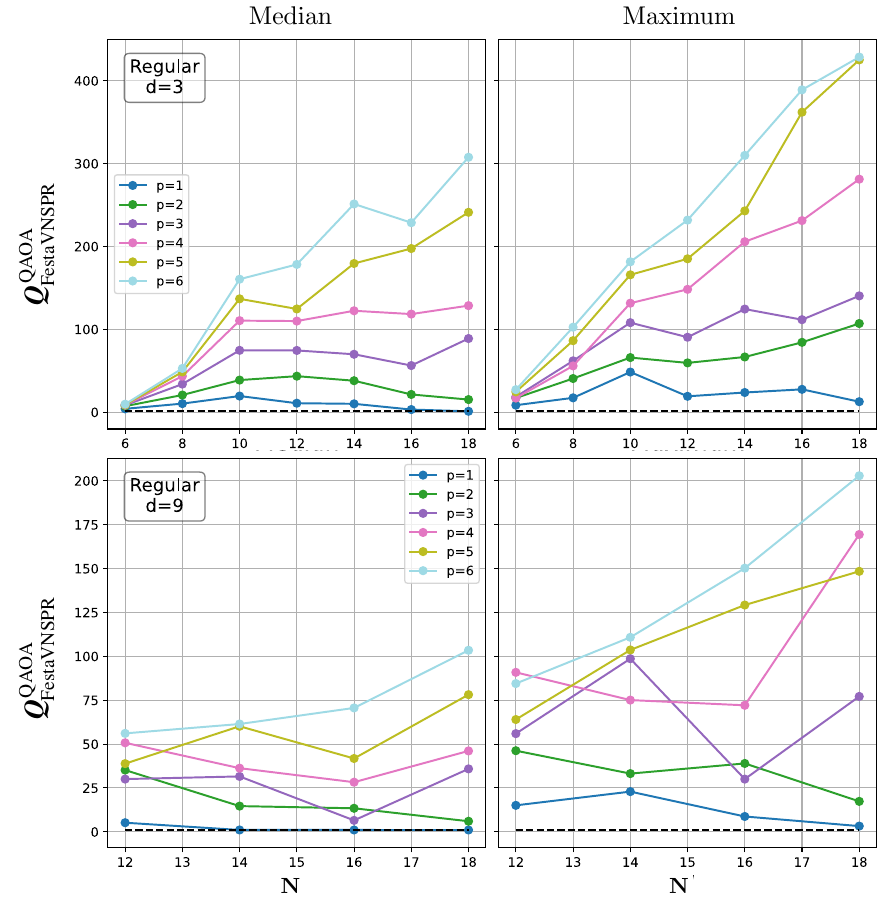}
\caption{The Q-factor from warm-starting the the FESTA2002VNSPR algorithm implemented in MQLib~\cite{dunning_what_2018} for unweighted Max-Cut with samples obtained using classical emulation of the QAOA algorithm using the `balanced' approach for predicting angles for 10 randomly generated random regular graphs with degree $d=3$ (top) and 10 graphs with degree $d=9$ (bottom). As in Figure \ref{fig:festa_er_different_p}, we see a much better performance in both the average and maximum cases for less dense graphs, where for $p=6$ layers we can achieve median Q-factors of over 300 for graphs with $d=3$, whereas just over to 100 for graphs with $d=9$.}
\label{fig:festa_regular_different_d}
\end{figure}

\begin{figure}[t]
\centering
    \includegraphics[width=0.9\linewidth]{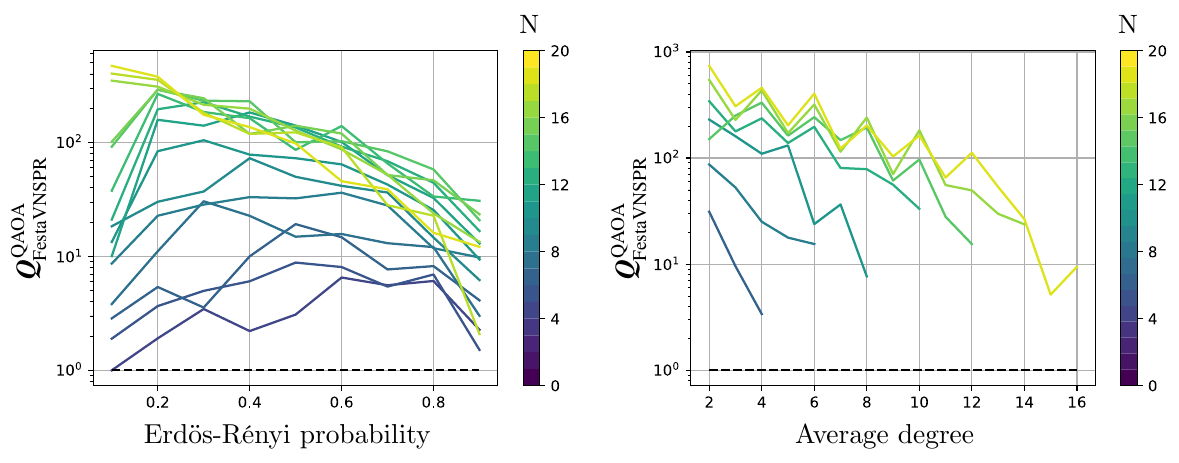}
\caption{The Q-factor from warm-starting the the FESTA2002VNSPR algorithm using classically emulated $p=6$ QAOA with the same set of Erd\H{o}s-R\'enyi graphs (left) and regular graphs (right) as Figure \ref{fig:emulated_random_graph_speedups_festa}, as a function of the density of the graph indicated by the probability of an edge and the average degree respectively. Each point averages over 10 graphs for unweighted Max-Cut. We see a very similar to behaviour to Figure \ref{fig:palubeckis_density_speedup}, where as the graph gets more dense, the Q-factor decreases.}
\label{fig:festa_vnspr_density_speedup}
\end{figure}

As well as using our method on the PalubeckisMST1 algorithm~\cite{palubeckis_multistart_2004} shown above and in in Section \ref{sec:results}, we also warm-started the FESTA2002VNSPR algorithm~\cite{festa_2002} implemented in MQLib~\cite{dunning_what_2018}. This heuristic is built from a base local search algorithm, with the enhancements of variable neighbourhood search (VNS) and path relinking (PR), which are metaheuristics that aim to improve the bit-strings that are found in the local search. We used the same set of graphs and parameters as those for PalubeckisMST1, and compute the Q-factor $Q^{QAOA}_{FESTA\_VNSPR}$. In Figure \ref{fig:emulated_random_graph_speedups_festa} we see a greater speedup as we increase the number of vertices and the number of QAOA layers, similar to the results in Figure \ref{fig:emulated_random_graph_speedups_palubeckis}. The quantitiative values of the Q-factor are higher for FESTA2002VNSPR compared to PalubeckisMST1, which may be because the former uses a greater number of shorter iterations, so a warm start will be able to cause the number of iterations to decrease more. Further experiments will need to be performed to determine the effect on the actual runtime of the algorithms as opposed to the number of iterations. We see a very similar picture for how the density of the graph affects the Q-factor in Figures \ref{fig:festa_er_different_p}, \ref{fig:festa_regular_different_d} and \ref{fig:festa_vnspr_density_speedup}. Again the denser the graph is, the smaller and more varied speedup we see over the problem instances. There is a larger gap between the median and maximum Q-factors in the denser graphs, indicating the performance is more dependent on the problem instance in these cases. The increase in Q-factor as the problem size increases is also higher for less dense graphs, indicating the warm start is helping more for these instances.

\clearpage

\section{Circuit Optimization}

\subsection{Description of A* Algorithms}\label{sec:a_star_description}

In general, graph search methods take a directed weighted graph $S(V, E, W)$ describing the state space, a starting node $s \in V$ and set of target nodes $T \subseteq V$, and output a path from $s$ to any $t \in T$ with small total weight. Due to the typically exponential size of the graph $S$, many graph search algorithms will explore this state space locally, starting at $s$ and iteratively picking neighbours until a node in $T$ is found. This requires that the neighbours of a given node are easily computable as well, as at every step of the algorithm we need to determine where it can explore next.

The A* algorithm~\cite{hart_a_star} is a heuristic algorithm that explores this graph $S$ one node at a time, maintaining a \emph{frontier} of possible nodes which could be explored next. Every time a node $n \in V$ is explored, it is removed from the frontier and all $n' \in V$ such that $(n, n') \in E$ are added to the frontier, unless $n \in T$ at which point the algorithm terminates. The choice of the next node to explore in the frontier is dependent on two factors, the current cost of the path to the current node $g: S \to \mathbb{R}$, and an estimate for the cost required to reach the target set $T$, $h: S \to \mathbb{R}$. More generally the cost can also depend on the path taken from $s$ to $n$, but as $S$ in our case is a tree, it suffices to define it to only depend on the node $n$. For a given frontier, the A* algorithm will choose the node $n$ such that $f(n) = g(n) + h(n)$ is minimised, so $f$ is a heuristic estimate for the length of the shortest path from $s$ to any $t \in T$ that passes through $n$.

The choice of cost function and heuristic has effects on the optimality and runtime of the algorithm. For example, by taking $h \equiv 0$ we get \emph{uniform cost search}, where the node that will be explored will always be the one with the smallest current cost. This guarantees optimality, but without a heuristic to guide the search it can potentially run very slowly. The other extreme is taking $g \equiv 0$, called \emph{greedy search}, where we only use the heuristic to guide the search. This is likely to be suboptimal, but good heuristics will mean it can find valid paths to $T$ very quickly.

The optimal heuristic $h^*$ is the heuristic that precisely is the length of the shortest path from a node $n$ to $T$. This is often intractable to compute exactly and so little use in practice, but any heuristic satisfying $0 \leq h \leq h^*$ is called \emph{admissible} and when used in the A* algorithm will lead to an optimal path \cite{russel_ai}. Furthermore, the closer $h$ and $h^*$ are, the faster the algorithm will find the optimal path. Despite this, non-admissible heuristics can still be effective at finding close to optimal paths much faster.

\begin{figure}[t]
\centering
    \includegraphics[width=\linewidth]{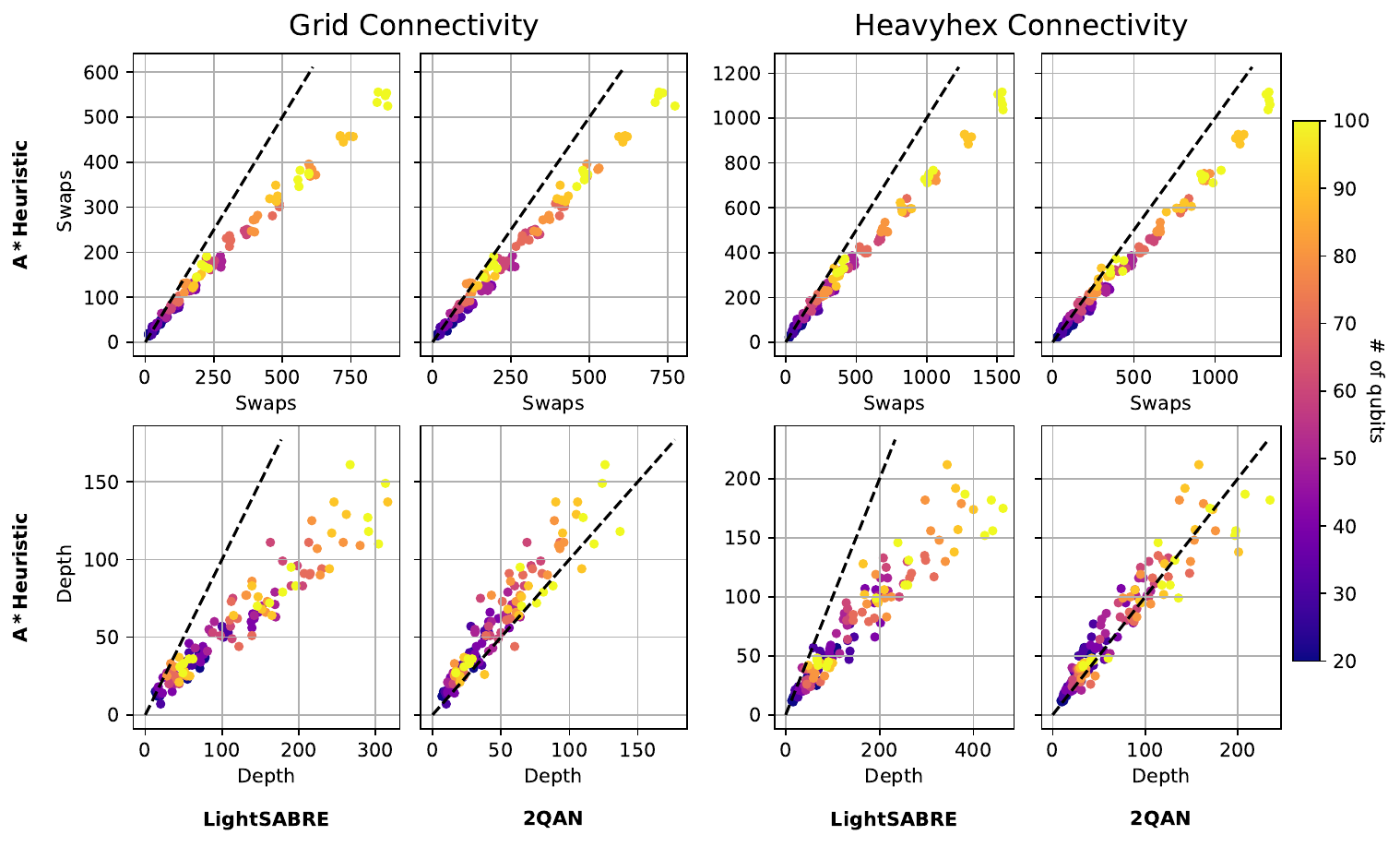}
\caption{The number of swaps required and depth required to implement the QAOA circuit for Max-Cut on random \{3, 5, 7\}-regular graphs with between $20$ and $100$ nodes on a $(12\times12)$-grid connectivity (left) and 156 qubit heavyhex connectivity on the IBM Heron chip (right), computed using the A* heuristic, 2QAN and LightSABRE. The A* heuristic consistently produces circuits with fewer swaps in both the grid and heavyhex connectivities against both other algorithms, and also produces shallower circuits compared to LightSABRE. However it often produces deeper circuits compared to 2QAN.}
\label{fig:further_routing_comparisons}
\end{figure}

\subsection{Further Circuit Optimization Results}\label{sec:appendix_further_routing}

In Section \ref{sec:circuit_optimisation_results}, we see that we get the best performance using the quadratic assignment vertex-to-qubit mapping and the A* heuristic routing algorithms, iterating the routing process 10 times using the final mapping of a previous iteration as the initial mapping of the next. Throughout we used the total distance heuristic with $q=1$. We compared how this performs against the existing mapping and routing software tools 2QAN~\cite{lao22}, and LightSABRE~\cite{zou24_sabre}. Here we ran the three methods on a wider class of problem instances, optimizing QAOA circuits for Max-Cut on five random regular graphs each with sizes $n \in \{20, 30, ..., 100\}$ and degrees $d = \{3, 5, 7\}$. As before, we transpiled these circuits for both the $(12 \times 12)$-grid connectivity and the IBM Heron heavyhex connectivity on 156 qubits. For the 2QAN comparison, we used the \verb|run_qiskit| initial mapping method implemented in that package. For the comparison with LightSABRE, we used the \texttt{lookahead} heuristic and 200 trials.

We compare the number of swaps required and depth of the resulting transpiled circuits using these three methods in Figure \ref{fig:further_routing_comparisons}. The A* heuristic consistently produces shallower circuits with fewer swaps compared to LightSABRE, requiring on average 23.2\% and 20.7\% fewer swaps and constructing circuits with depths 42.7\% and 42.7\% smaller in the grid and heavyhex connectivities respectively. The largest reduction in the number of swaps was 40.7\% and the largest decrease in the depth was 65.3\%. Compared to 2QAN it consistently produces a circuit with a smaller number of swaps, requiring on average 18.7\% and 19.5\% fewer swaps in the grid and heavyhex connectivities respectively. However, the depth on average increased by 31.1\% and 15.2\% in the grid and heavyhex connectivities respectively. This is likely because the cost function $g$ used in the A* heuristic algorithm is the number of swaps, and independent of the depth. This means it will not necessarily pick the less deep circuit given two circuits with the same number of swaps at any given point in the algorithm, and furthermore may even prioritise decreasing the number of swaps at the expense of having a deeper circuit. Therefore if the main error source when running a circuit is the gate error, then the A* heuristic consistently produces better circuits, but if the aim of the transpilation technique is to reduce the depth of the circuit, then 2QAN may perform better. However, there are other cost functions $g$ and heuristics $h$ that can be used in the A* algorithm, and by introducing the depth into either of these will change what is prioritises during the search and may result in circuits that have smaller depths. Further work is required to determine what the best cost functions and heuristics are optimal for different purposes.

\begin{figure}[h!]
    \includegraphics[width=\textwidth]{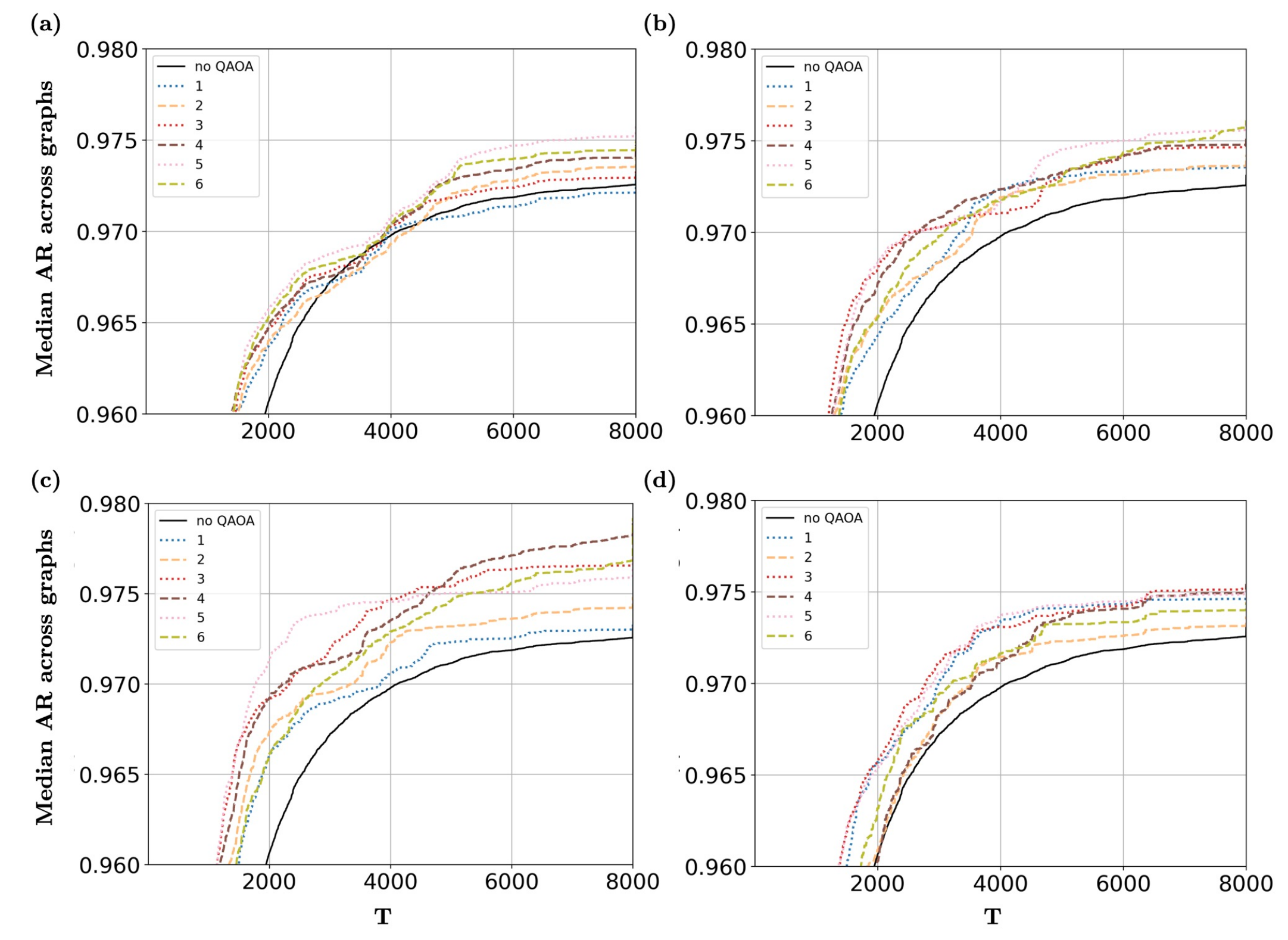}
    \caption{Median approximation ratio (Eq.~\eqref{eq:approx_ratio}) across all enhanced IBM graphs of size $41$ and across $1000$ optimizations each using either QAOA or random initial bit-strings for the PalubeckisMST1 algorithm. The random initialization results are plotted in black. We compare the cases with (a) just using the QAOA samples, (b) using only the $1000$ lowest energy samples (out of $10000$), (c) using samples with the Hamming filter applied and (d) using samples with frequency filtering applied. Note that in all cases, the hardware samples include LNF readout error mitigation, as described in the main text.}
    \label{fig:ibm_maxcut_41_vertex_approx_ratio}
\end{figure}

\section{Further Hardware Results on SWAP-Enhanced 41 Vertex Graphs}
\label{sec:appendix_further_hardware}

Supplementing the results in the main text (Sec.~\ref{sec:enhanced_ibm_hw_results}), in Fig.~\ref{fig:ibm_maxcut_41_vertex_approx_ratio} we plot the median approximation ratio obtained across 1000 optimizations with PalubeckisMST1 using either random bit-strings or QAOA warm starts obtained from hardware. We compare the outcomes for different post-processing filtering techniques and with the LNF readout error mitigation applied.

The approximation ratio for a graph instance at $T$ iterations is defined as:
\begin{align}\label{eq:approx_ratio}
    \text{AR} = \frac{C^{\rm Max-Cut}(T)}{C^{\rm Max-Cut}_{\rm opt}}
\end{align}
where $C^{\rm Max-Cut}(T)$ is the value of the Max-Cut cost function, here treated as a maximization of the cut value, obtained by the optimizer after $T$ iterations and $C^{\rm Max-Cut}_{\rm opt}$ is the true optimal value. 

\begin{figure}[t]
    \includegraphics[width=\textwidth]{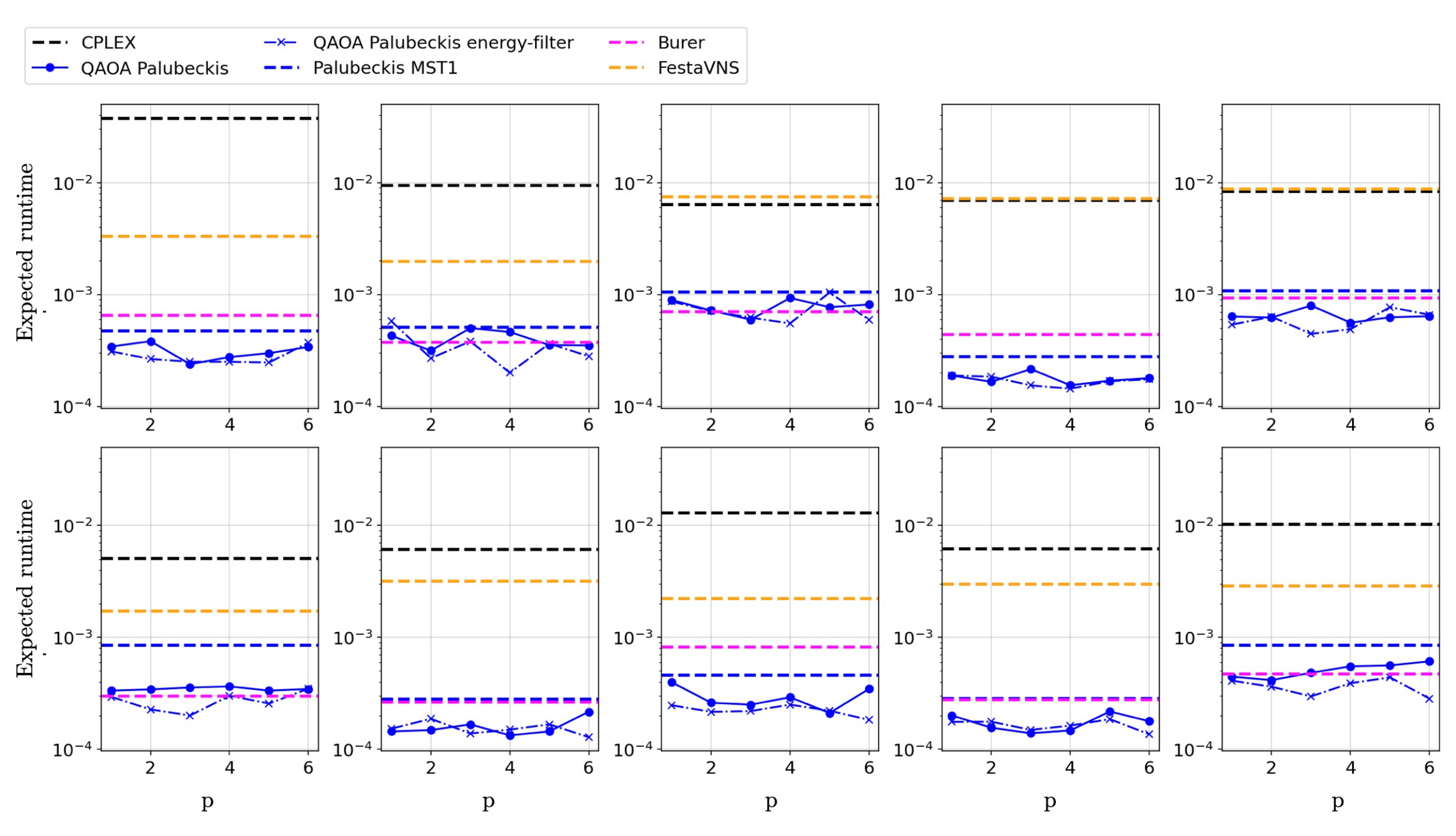}
    \caption{Comparing the expected runtimes of several of the best heuristics described in~\cite{dunning_what_2018} as well as CPLEX~\cite{cplex2009v12} to the QAOA warm-started version of PalubeckisMST1. Each plot corresponds to a graph instance of size $41$ as described in Sec.~\ref{sec:enhanced_ibm_hw_results}. The QAOA bit-strings used for the experiments are post-processed with the energy and the Hamming filter. The horizontal dashed lines correspond to the expected runtimes (Eq.~\eqref{eq:expected_runtime}) of each of the optimization algorithms without any modifications from their original implementation, while the circle and cross plots correspond to expected runtimes when initialising Palubeckis with QAOA samples obtained from a run with $p$ QAOA layers (without and with energy filtering, respectively. Note that these plots are comparing expected runtimes without the time required to generate/obtain the resource bit-strings for each case.}
    \label{fig:ibm_41graph_runtimes}
\end{figure}

We also plot the individual expected expected runtime estimates for the $10$ separate $41$-vertex enhanced IBM graphs in Fig.~\ref{fig:ibm_41graph_runtimes}. These were used to compute the median values in Fig.~\ref{fig:ibm_runtimes_combined} and allow for a more detailed insight into the behaviour of classical and quantum-enhanced algorithms across different problem structures.

\section{Classical benchmarks}\label{sec:classical_benchmarks}

The performance of local search heuristics is often difficult to assess, as they may explore and exploit the solution landscape in radically different ways. This is exemplified by the fact that Palubeckis~\cite{palubeckis_multistart_2004}, one of the main algorithms we investigate in this work, does not trivially solve Max-Cut on line graphs (see discussion in Sec.~\ref{sec:hardware}). For some classes of problem instances, one heuristic may perform exceedingly well, while another might fail entirely due to its method of exploring the search space.

\begin{figure}
    \includegraphics[width=\textwidth]{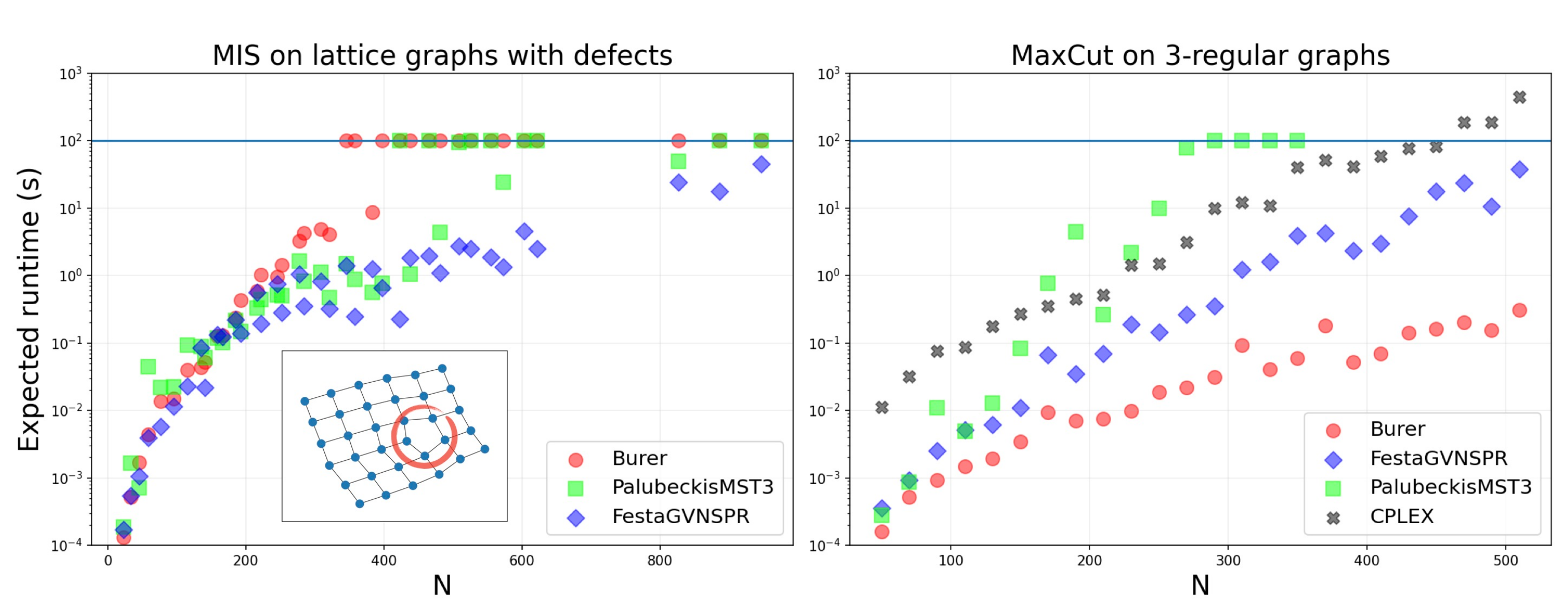}
    \caption{Plot of the expected runtimes (Eq.~\eqref{eq:expected_runtime}) of several of the best ranked algorithms from Ref.~\cite{dunning_what_2018}. The plot on the left is for solving MIS on lattice graphs of different dimensions with added defects as described in the text (see example of defect in inset figure) while the plot on the right is the Max-Cut problem on random $3$-regular graphs. The runtime estimates for the heuristics are obtained by running an optimization on each problem instance $50$ times with a time limit of $2$ seconds, meaning the `time-out' limit for all runs is $100$ seconds, as indicated on the plot. For CPLEX, the time-out is set to $5000$s. For each problem size $N$ of the random regular Max-Cut plots, the data point is obtained as the median across $5$ random problem instances.} 
    \label{fig:classical_benchmarks}
\end{figure}

In Fig.~\ref{fig:classical_benchmarks} we plot the expected runtimes of some of the best-performing algorithms from Ref.~\cite{dunning_what_2018}: Burer~\cite{burer_rank-two_2002}, FestaGVNSPR~\cite{festa_2002} and PalubeckisMST3~\cite{palubeckis_multistart_2004}, as well as CPLEX~\cite{cplex2009v12}, which was used to exactly determine the optimal value for each of the graphs above. Burer and Festa are Max-Cut-focused algorithms, while Palubeckis was developed for general QUBOs. We test the algorithms on two different structures of problems: unweighted Max-Cut on $3$-regular graphs and MIS on lattice graphs with added defects. The defects are inserted into the lattices by creating domain walls in the MIS solution, which on a lattice would just be a checkerboard pattern, by removing a node in the middle of the lattice and contracting the row which follows the node. For an example of such a defect on a small lattice, see inset in left plot of Fig.~\ref{fig:classical_benchmarks}. We choose this problem type as while it adds difficulty for local algorithms due to the non-local impact a domain wall can have on the solution, it is easy to verify the optimality of a solution once it is found due to the structure of the graphs. In general, it is difficult to benchmark larger graphs without such structure, as it is difficult to verify the optimal solution without relying on exact solvers, whose runtime often scales very poorly.

Our results show that while in the $3$-regular graph case, the scaling of Burer is clearly better than either Festa or Palubeckis, the opposite is true in the case where MIS on lattice graphs with defects is considered. Festa manages to solve graphs with just under $1000$ nodes in less than $100$ seconds, while Burer times out at around $400$ nodes. We note that we could not benchmark random-regular graphs of a higher degree for reasonable problem sizes as CPLEX would time out, thus giving us no optimal value to compare against. 

\end{document}